\documentclass[aps,preprint,superscriptaddress,showpacs]{revtex4}
\usepackage[english]{babel}
\usepackage{epsfig}
\usepackage{graphics}
\usepackage{natbib}
\usepackage{amsthm}
\usepackage{amsmath}
\usepackage{times}
\usepackage{psfrag}
\usepackage{subfigure}
\usepackage{epstopdf}
\begin{document}

\title{L\'evy noise induced transitions and enhanced stability  in  a birhythmic van der Pol system}

\author{\textbf{Ren\'e Yamapi}}
\email[Author to whom correspondence should be addressed. Electronic mail : ] {ryamapi@yahoo.fr}
\affiliation{Fundamental Physics Laboratory, Physics of Complex System group,
Department of Physics, Faculty of
 Science, University of Douala, Box 24 157 Douala, Cameroon.}
\affiliation{Potsdam Institute for Climate Impact Research (PIK), 14473 Potsdam, Germany}

\author{\textbf{Raoul Mbakob Yonkeu}}
\affiliation{Laboratory of Mechanics and Materials, Department of Physics, Faculty of
Science, University of Yaound\'e I, Box 812, Yaound\'e, Cameroon.}
\author{\textbf{Giovanni Filatrella}}
 \affiliation{Department  of Sciences and Technologies
\small and Salerno unit of CNSIM, University of Sannio, Via Port'Arsa 11,
I-82100 Benevento, Italy.}

\author{\textbf{J\"urgen Kurths}}
\affiliation{Potsdam Institute for Climate Impact Research (PIK), 14473 Potsdam, Germany}
\affiliation{Department of Physics, Humboldt University, 12489 Berlin, Germany.}

%\author{Also possible collaborators: C. Guarcello, V. Pierro}

\date{\today}

\begin{abstract}
This work describes the effects of L\'evy noise on a birhythmic van der Pol like oscillator.
Numerical simulations demonstrate that the noise induced escapes from an
attractor to another are not markedly different from escapes between
stable points in an ordinary potential, albeit the attractors are
separated by a barrier of a quasi (or pseudo) potential.
However, some differences appear, and are more pronounced when
the  L\'evy distribution index is close to two.
\end{abstract}
\maketitle

%\newpage

\section{Introduction}
\label{introduction}

L\'evy noise is a preeminent example of non Gaussian noise.
The underlying notion is that some noise sources or random signals
are not characterized by a finite variance.
Moreover, if one makes the further requirement that the noise
distribution is stable, that is, it is the limit distribution
of the sum of many random and identically distributed variables,
the resulting distribution belongs to the family of L\'evy functions,
as a generalization of the central limit theorem \cite{Dubkov08}.
The presence of anomalous, that is non thermal (Gaussian) noise is
widely recognized in physical systems and practical devices.
In the natural world, L\'evy noise has been modeled for
most different systems, from predator-pray \cite{LaCognata10}.
Quite naturally, L\'evy noise often appears in telecommunications
and networks \cite{Yang03} where  noise (from diverse sources
as atmospheric disturbances, relay contacts, electromagnetic
devices, electronic apparatus,  transportation systems,
switching transients, and accidental hits in telephone
lines \cite{Bhat06}) can exhibit impulsive and L\'evy-type characteristics.
In mechanical systems, L\'evy fluctuations have been also
used to describe vibration data in industrial bearings \cite{Li10,Chou14,Saad15}.
Another example of the relevance of anomalous distributions
in material issues has been proposed in photoluminescence
experiments of moderately doped n-InP samples \cite{Lury12},
and the presence of L\'evy flights could prove to have an impact
on the design of some optoelectronic devices  \cite{Suba14}.
On the fundamental side L\'evy processes can reveal the properties
in the electron transport \cite{Novi05} and optical properties
\cite{Kuno01,Mess01,Brok03} of semiconducting nanocrystals quantum dots.
Transport properties can be connected with L\'evy superdiffusion \cite{Metzler00},
\emph{e.g.} in the quasiballistic heat conduction \cite{Verm15,Moha15}.

A special role is played by L\'evy noise in oscillators. Most
important for the electrical power grid is the noise from wind
turbines rotation parts \cite{Elya16}, that can severely affect the infrastructure stability.
Also in superconducting active oscillators as superconducting Josephson
junctions \cite{Barone82}: L\'evy noise has been advocated, for example,
in graphene based devices in the form of rare jumps of the voltage
response of such non linear oscillator  \cite{Guarcello16} or when
the electron-electron interaction of the graphene under the effects
of a laser source, gives rise to a random walk with L\'evy flights distribution \cite{Bri14}.
In nonlinear oscillators the effect of L\'evy noise, and indeed of
Gaussian noise as well, is subtle, in the sense that the stable solution is a dynamical attractor.
When this is the case, and the force cannot be derived as the gradient
of a potential function, the problem of the stability cannot be
deduced from the standard escape from a potential well, characterized
by an \emph{Arrhenius} behavior of the lifetime (as a function of the noise intensity).
However, an alternative approach based on the concept an effective
potential (a pseudo, or quasi-potential) has proved effective to
treat the consequences of noise on the metastable periodic attractor
for nonlinear oscillators as Josephson junctions \cite{Kautz88,Kautz94,Yamapi14}
and for van der Pol birhythmic oscillators \cite{Yamapi10,Yamapi12,Mbakob2},
also in the presence of correlated noise \cite{mbakob3}.
It is therefore quite natural to imagine to extend the theory of L\'evy noise
induced escapes from ordinary potentials \cite{chechkin1,chechkin2} to quasi-potentials for birhythmic van der Pol like systems.
In doing so, one follows in the essence the approachh
(for monorhythmic systems) already employed using the principle
of minimum action \cite{Sun17,Wan18}, that describes a numerical
method to derive the quasi-potential for a non gradient system.
The conceptual difficulty to follow this line of reasoning is that
the distance rather than the energy barrier matters. Moreover, this line of research is numerically heavy.
An analytic approach has been presented for the ordinary,
non birhythmic, van der Pol system (with L\'evy noise)  \cite{Hogele14},
and also extended to a birhythmic system \cite{Moran84}.

In this work we use the numerical method already employed for Gaussian noise, that is to revert the logic of the escape time in the presence of L\'evy noise \cite{Chechkin05,Chechkin07}.
In doing so one wishes to determine if the quasi-potential
concept is applicable to birhythmic system, and the limits of the applicability of such concept.
Also, one wishes to ascertain if numerical simulations can carry
information about the theoretical estimates of the features
of escapes from ordinary potentials, for instance the
dependence of the escape times on the L\'evy index, $a$.

The paper is organized as follows. Sect.II describes the birhythmic van der Pol  system driven by L\'evy noise and the algorithm of the numerical simulations.
After the description of the main features of the deterministic birhythmic van der Pol oscillator and the parameter region where birhytmicity appears.
The L\'evy's process and numerical algorithm conclude this section.
In Sect. III, we focus on numerical computed  escape rates and the algorithm to generate random L\'evy noise and to integrate the stochastic differential equation.
For low noise regime, the Arrhenius factor (\emph{i.e.}, the relation between the escape time $T_{esc}$ and the noise intensity $D$) allows to determine an effective activation energy barrier $\Delta U$ from the slope of the linear part of the variation in the
escape time versus the inverse noise intensity.
 Section IV is devoted to conclusions.

\section{The stochastic birhythmic system and Levy's process}
\subsection{The birhythmic van der Pol system}

The model considered is a  van der Pol-like oscillator with a nonlinear function of higher polynomial order described by the following nonlinear equation (overdots as usual stand for the derivative with respect to time)
\begin{equation}\label{eq1}
  \ddot x-\mu\left(1-x^2+\alpha x^4-\beta x^6\right)\dot x+x=0,
\end{equation}
where the quantities $\alpha$ and $\beta$ are positive parameters indicate
the system behavior to a ferroelectric instability compared
with its electrical resistance, while $\mu$ is a positive parameter that tunes nonlinearity \cite{kaiserv,kaiserp,enjieu-yamapi-chabi,enjieu-chabi-yamapi-woafo}.
 Eq.~(\ref{eq1}) describes several dynamic
systems, ranging from physics to engineering and biochemistry
\cite{enjieu-chabi-yamapi-woafo,yamapi-nana-enjieu-2007,vanderpolgeneral}.
In particular Eq.~(\ref{eq1}) seems to be more
appropriate for some biological processes than the classical van
der Pol oscillator, as shown by Kaiser in Ref. \cite{kaiser2}.
When employed to model biochemical systems, namely
the enzymatic-substrate reactions, $x$ in Eq.~(\ref{eq1})
is proportional to the population of enzyme
molecules in the excited polar state.
Model (\ref{eq1}) is therefore a prototype for self-sustained systems and exhibits some interesting features
of nonlinear dynamical systems; for instance Ref. \cite{kaiserv,kaiserp} have analyzed the super-harmonic
resonance structure and have found symmetry-breaking crisis and
intermittency.
The nonlinear dynamics and the synchronization
process of two such systems have been recently investigated in
Ref.~\cite{enjieu-chabi-yamapi-woafo,enjieu-yamapi-chabi}, while the possibility that
introducing an active control of chaos can be tamed for an appropriate
choice of the coupling parameters has been considered in Ref.~ \cite{yamapi-nana-enjieu-2007}.
Recently, we have found in ref.~\cite{mbakob3} the effects of external
excitation on the multi-limt cycle van der Pol system.
It appears that the birhythmic behavior is still present for a very small
excitation and disappear when the amplitude of the driven  becomes large.

The nonlinear self-sustained oscillator Eq.~(\ref{eq1}) possesses more than one stable limit-cycle solution
\cite{kaiser2}, a condition for the occurrence of birhythmicity.
Birhythmic systems are of interest, for example in biology, to describe
the coexistence of two stable oscillatory states, a situation that
can be found in some enzyme reactions \cite{li}. Another example
is the explanation of the existence of multiple frequency and
intensity windows in the reaction of biological systems when they
are irradiated with very weak electromagnetic fields
\cite{kaiserp,kaiser2,kaiser3,kaiser4,kaiser5,kaiser6}.  In this work we
will focus on model (\ref{eq1}) as a prototype for the occurrence
of birhythmicity.

\begin{table}
  \begin{center}
  \begin{scriptsize}
  \begin{tabular}{|c|c|c|}
  \hline
  $AS_i=(a,\beta)$ & Amplitudes & Quasi-potential barriers $\Delta U_i$ \\
  \hline
  $AS_1=(0.114,0.003)$ &
  \begin{tabular}{c} $A_1=2.37720$ \\ $A_2=5.02638$ \\ $A_3=5.46665$ \end{tabular}
  & \begin{tabular}{c} $\Delta U_1 = 1.062 \times 10^{-2}$ \\ $\Delta U_3=2.111 \times 10^{-4}$ \end{tabular} \\
  \hline $AS_2=(0.1,0.002)$ &\begin{tabular}{c} $A_1=2.3069$ \\ $A_2=4.8472$ \\ $A_3=7.1541$ \end{tabular}
  & \begin{tabular}{c} $\Delta U_1 =1.437 \times 10^{-2} $ \\ $\Delta U_3=4.423 \times 10^{-2}$ \end{tabular}  \\
  \hline $AS_3=(0.12,0.003)$ & \begin{tabular}{c} $A_1=2.4269$ \\ $A_2=4.2556$\\ $A_3=6.3245$  \end{tabular}
    & \begin{tabular}{c} $\Delta U_1 = 4.989 \times 10^{-2}$ \\ $\Delta U_3=2.372 \times 10^{-2}$ \end{tabular}    \\
  \hline $AS_4=(0.13,0.004)$ & \begin{tabular} {c} $A_1=2.4903$ \\ $A_2=4.4721$ \\ $A_3=5.0791$ \end{tabular}
    & \begin{tabular}{c} $\Delta U_1 = 4.342 \times 10^{-3}$ \\ $\Delta U_3=4.237 \times 10^{-4}$ \end{tabular} \\
  \hline
  \end{tabular}
  \end{scriptsize} %scriptsize
  \caption{\it{Amplitudes of the limit cycles and quasi-potential barriers $\Delta U_{1,3}$
  for the asymmetric potential. All data refer to the case $\mu=0.01$ and $D=0.0$.
  All the results are obtained analytically.}}
  \label{Tab1}
  \end{center}
\end{table}

\begin{table}
  \begin{center}
  \begin{scriptsize}
  \begin{tabular}{|c|c|c|}
  \hline
  $S_i=(a,\beta)$ & Amplitudes & Quasi-potential barriers $\Delta U_i$\\
  \hline $S_1=(0.0675,0.0009)$ & \begin{tabular}{c} $A_1=2.1730001$ \\ $A_2=6.3245003$ \\ $A_3=8.6760004$ \end{tabular} & \begin{tabular}{c} $\Delta U_1 = 6.5085\times 10^{-2}$ \\ $\Delta U_3= 6.5085\times 10^{-2}$ \end{tabular} \\
  \hline $S_2=(0.1635,0.007)$ & \begin{tabular} {c} $A_1=3.0925001$ \\ $A_2=3.5280002$ \\ $A_3=3.9190002$ \end{tabular}
  & \begin{tabular}{c} $\Delta U_1 =  2.376\times 10^{-5}$ \\ $\Delta U_3= 2.445\times 10^{-5}$ \end{tabular} \\
  \hline $S_3=(0.16,0.00658)$ & \begin{tabular} {c} $A_1=2.9520001$ \\ $A_2=3.5965002$ \\ $A_3=4.1535002$ \end{tabular} & \begin{tabular}{c} $\Delta U_1 = 1.0345 \times 10^{-4}$ \\ $\Delta U_3= 1.1011\times 10^{-4}$ \end{tabular} \\
  \hline $S_4=(0.1476,0.0053)$ & \begin{tabular}{c} $A_1=2.6905001$ \\ $A_2=3.8525002$\\ $A_3=4.7405002$  \end{tabular} & \begin{tabular}{c} $\Delta U_1 =  8.6713\times 10^{-4}$ \\ $\Delta U_3= 8.7677\times 10^{-4}$ \end{tabular} \\
  \hline
  \end{tabular}
  \end{scriptsize} %scriptsize
  \caption{\it{Amplitudes of the limit cycles and quasi-potential
  barriers $\Delta U_{1,3}$ for the symmetric potential. All data refer to the case $\mu=0.01$
  and $D=0.0$.
  All the results are obtained analytically.}}
  \label{Tab2}
  \end{center}
\end{table}

\begin{figure}%[htbp]
\begin{center}
\includegraphics[height=6.4cm,width=12.4cm]{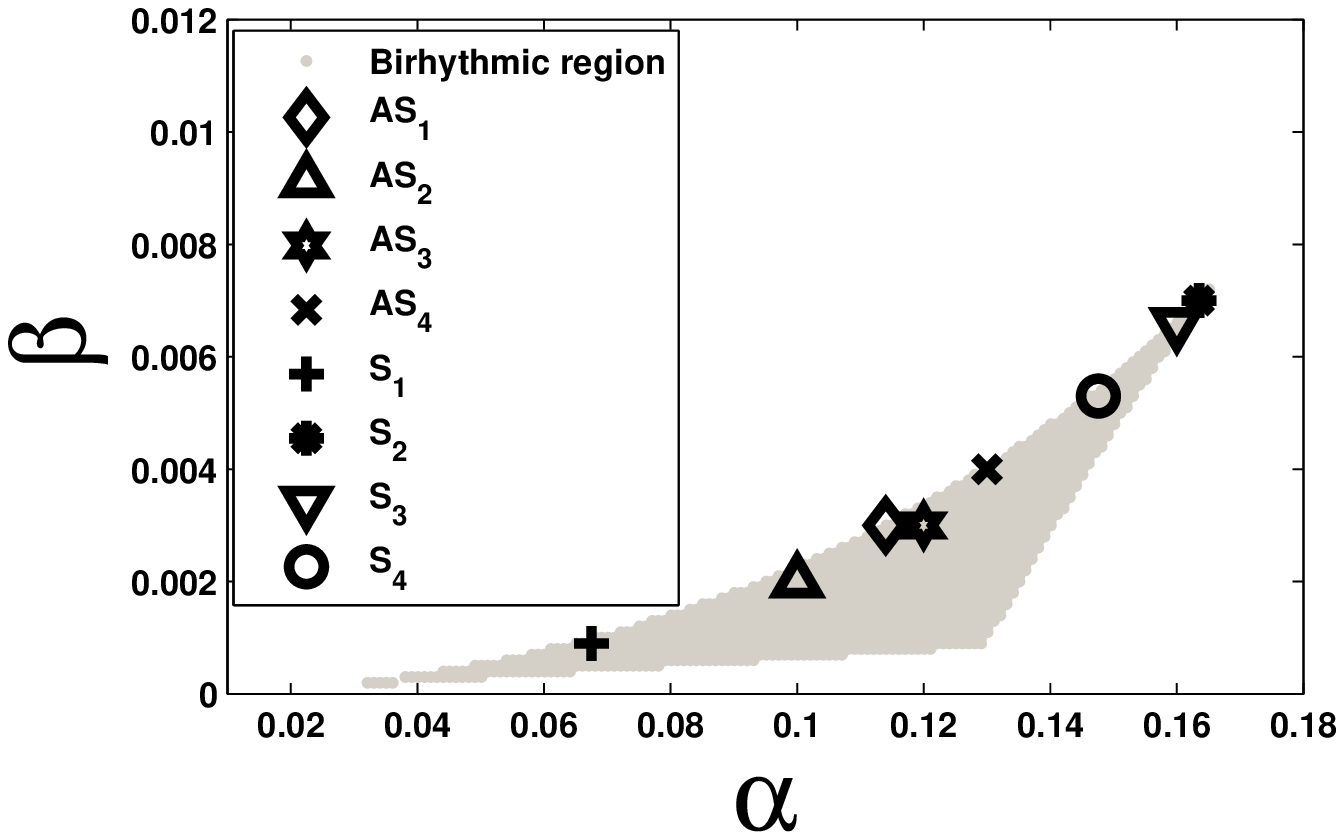}
\caption{\it
Parameter region of the single limit cycle (white area) and
three limit cycles (gray area) with $\mu=0.01$ as obtained from
simulations of Eq.(\ref{eq3}).
The symbols refer to the parameter sets investigated in this work, see Tables \ref{Tab1} and \ref{Tab2}.
The other parameter is $\mu=0.01$.}
\label{fig1}
\end{center}
\end{figure}

Following Refs.\cite{enjieu-chabi-yamapi-woafo,enjieu-yamapi-chabi}, the periodic solutions of Eq. (1) can be
approximated by
\begin{equation}\label{eq2}
  x(t)=A\cos \omega t.
\end{equation}
The analytic amplitude $A$ and frequency $\omega$ can be readily obtained \cite{enjieu-yamapi-chabi,enjieu-chabi-yamapi-woafo}.
It has been found that the amplitude $A$  is independent  of the  parameter $\mu$, which  only enters in the frequency $w$. The amplitude equation is given by
\begin{equation}\label{eq3}
  \frac{5\beta}{64}A^6-\frac{\alpha}{8}A^4+\frac{1}{4}A^2-1=0,
\end{equation}
and $\omega$ is given by:
\begin{eqnarray}
\label{eq4}
\omega=1+\mu^2\omega_2+o(\mu^3),
\end{eqnarray}
with
\begin{equation*}
\omega_2=\frac{93\beta^2}{65536}A^{12}-\frac{69\alpha\beta}{16384}A^{10}+(\frac{67\beta}{8192}+
\frac{3\alpha^2}{1024})A^8-(\frac{73\beta}{2048}+\frac{\alpha}{96})A^6+(\frac{1}{128}+
\frac{\alpha}{24})A^4-\frac{3}{64}A^2.
\end{equation*}
Depending on the value of the parameters $\alpha$ and
$\beta$, the van der Pol birhythmic system possesses one or  three limit cycles.
We find  that depending on the values of the parameters
$\alpha$ and $\beta$, the modified van der Pol equation (\ref{eq1}) possesses one or three limit cycles.
When three limit cycles are obtained, two of them are stable and one is unstable, a condition for
birhythmicity; the unstable limit cycle represents the separatrix
between the basins of attraction of the two stable limit
cycles. This appear in Fig. \ref{fig1} where the bifurcation lines that contour
the region of existence of birhythmicity in the two parameter
phase space $(a-\beta)$ \cite{enjieu-yamapi-chabi,enjieu-chabi-yamapi-woafo}.
 The bifurcation line on the left denotes the passage from a single limit cycle to three limit
cycles, while the right line denotes the reverse passage from
three limit cycles to a single solution. At the conjunction, a
codimension-two bifurcation, or cusp, appears. The first bifurcation
encountered increasing the amplitude $A$ corresponds to the
saddle-node bifurcation of the outer or larger limit cycle amplitude,
while the second bifurcation occurs in correspondence
of a saddle-node bifurcation of the inner or smaller amplitude
cycle.
The two frequencies associated with the limit cycles are very similar close to the lowest amplitude $A$ bifurcation and clearly distinct at the highest $a$ bifurcation line.
In Figure \ref{fig1} is is shown the range of existance of birhythmic solutions in the parameter plane $(a, \beta)$; examples of the corresponding pseudo-potential is shown in Figure \ref{fig2}:
 i) the first type is an asymmetric pseudo-potential with different potential wells (such as in Fig.\ref{fig2}(i));
 ii) and the second type is a symmetrical potential (see Fig.\ref{fig2}(ii)) and here the depths of the two potential wells are almost identical.
In Tables \ref{Tab1} and \ref{Tab2} are reported the amplitudes of the limit cycles in both cases,  for some selected values of $\alpha$ and $\beta$.
The two sets of parameters: $AS_1$ for the asymmetric quasi-potential and $S_1$ for the symmetric quasi-potential are considered in this work.

\subsection{The L\'evy process representation}

%The set of characteristic functions for the densities that generate the stable L\'evy processes.
L\'evy distributions, that describe the noise we consider in this work, are a rich class of probability distributions with several intriguing mathematical properties \cite{r1}.
The L\'evy process can be viewed as a generalized Wiener process that follow the L\'evy distribution $L_{a,b}(\zeta,\sigma,\delta)$; the representation is given by the characteristic
function defined in the Fourier transform $\Phi(k)$ \cite{r2}:
\begin{eqnarray}
\label{eq5}
  \Phi(k) =
  \left\{
    \begin{array}{lll}
      \exp\left\{ i\delta k-\sigma^a\mid k\mid^a(1-ibsgn(k)\tan(\frac{a\pi}{2}))\right\} & for &a\neq 1 ,2,\\
       \exp\left\{i\delta k-\sigma \mid k\mid(1+ib\frac{2}{\pi} sgn(k)\ln\mid k\mid)\right\}& for & a=1,\\
      \exp\left\{i\delta k-\frac{1}{2}\sigma^2k^2\right\} &for &a=2,\\
    \end{array}
  \right.
\end{eqnarray}
where $a$ $(0 < a \le 2)$ denotes the stability L\'evy index; for $a = 2$ the L\'evy stable distribution is the standard Gaussian distribution.
The parameter $b$ $(b \in[-1;+1])$ is an asymmetry, or skewness parameter, namely, the L\'evy distribution is symmetric for $b = 0.0$ and asymmetric for $b \neq 0.0, \delta$ $(\delta\in  \Re)$ is the center or location parameter which denotes the mean value of the
distribution, and the mean of the distribution exists and reads $\delta$ as $1 < a \leq 2$ \cite{r2}. The parameters $\sigma (\sigma\in  ]0;+\infty[)$ and $ D = \sigma^a$ are the scale parameter and noise intensity, respectively \cite{r3}.
The intensity of the L\'evy noise is determined by the parameter  $D$; if the van der Pol oscillator is used to describe the ferroelectric oscillations, $D$ has the physical meaning of the measure of the intensity of the random electric field.

\begin{figure}%[htbp]
\begin{center}
\includegraphics[height=5.0cm,width=16.4cm]{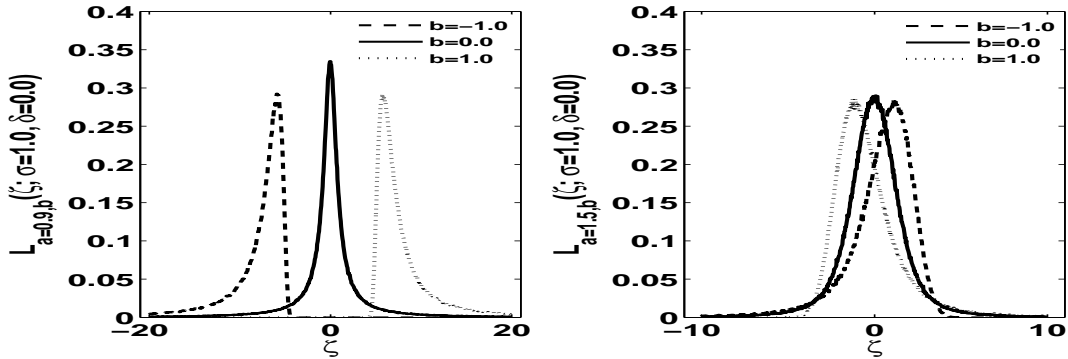}
\caption{\it
Sample a-stable probability density functions (PDF) with
$a = 0.9$(left panel) and $a = 1.5$ (right panel). For $b = 0$ distributions
are symmetric, while for $b \neq  0$ they are asymmetric functions.}
\label{fig1a}
\end{center}
\end{figure}

\begin{figure}%[htbp]
\begin{center}
\includegraphics[height=5.0cm,width=8cm]{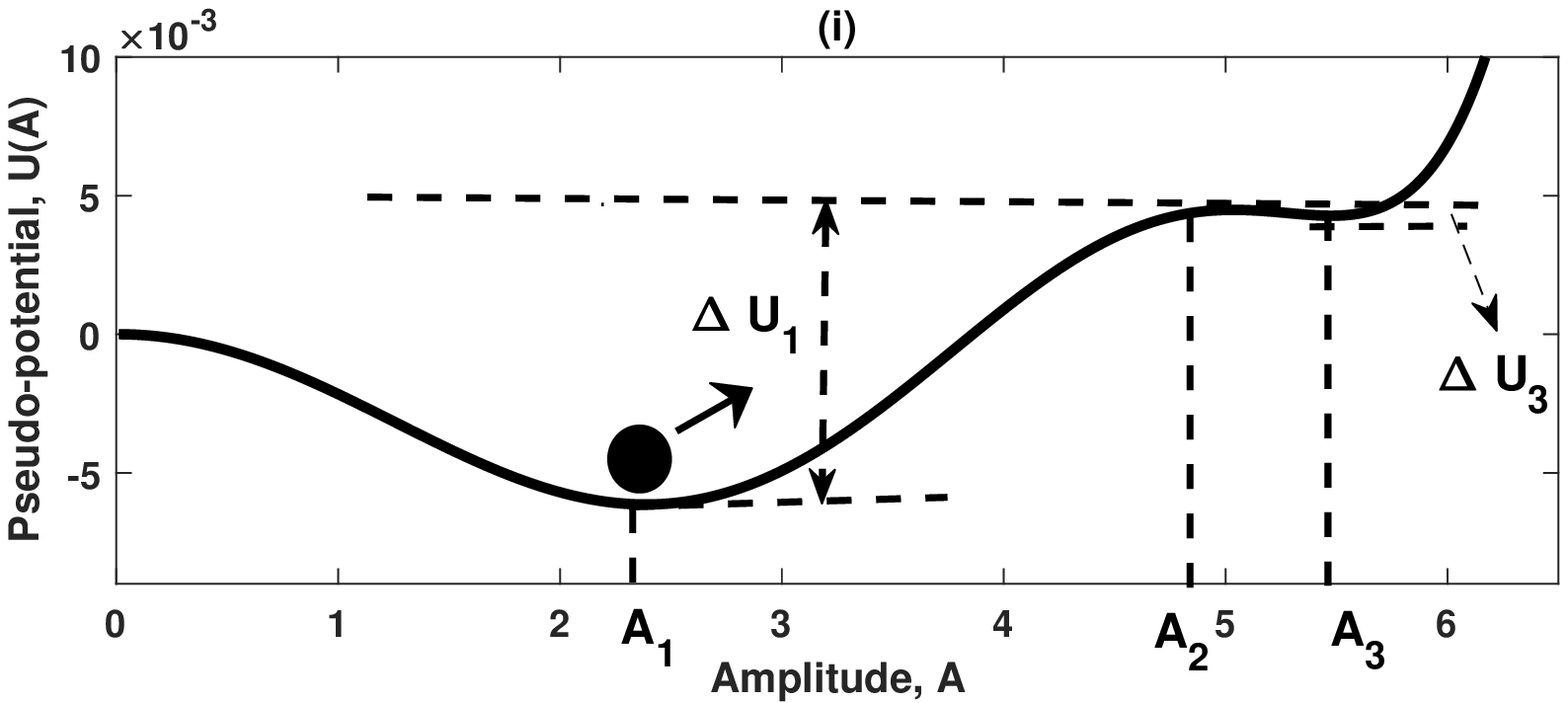}
\includegraphics[height=5.0cm,width=8cm]{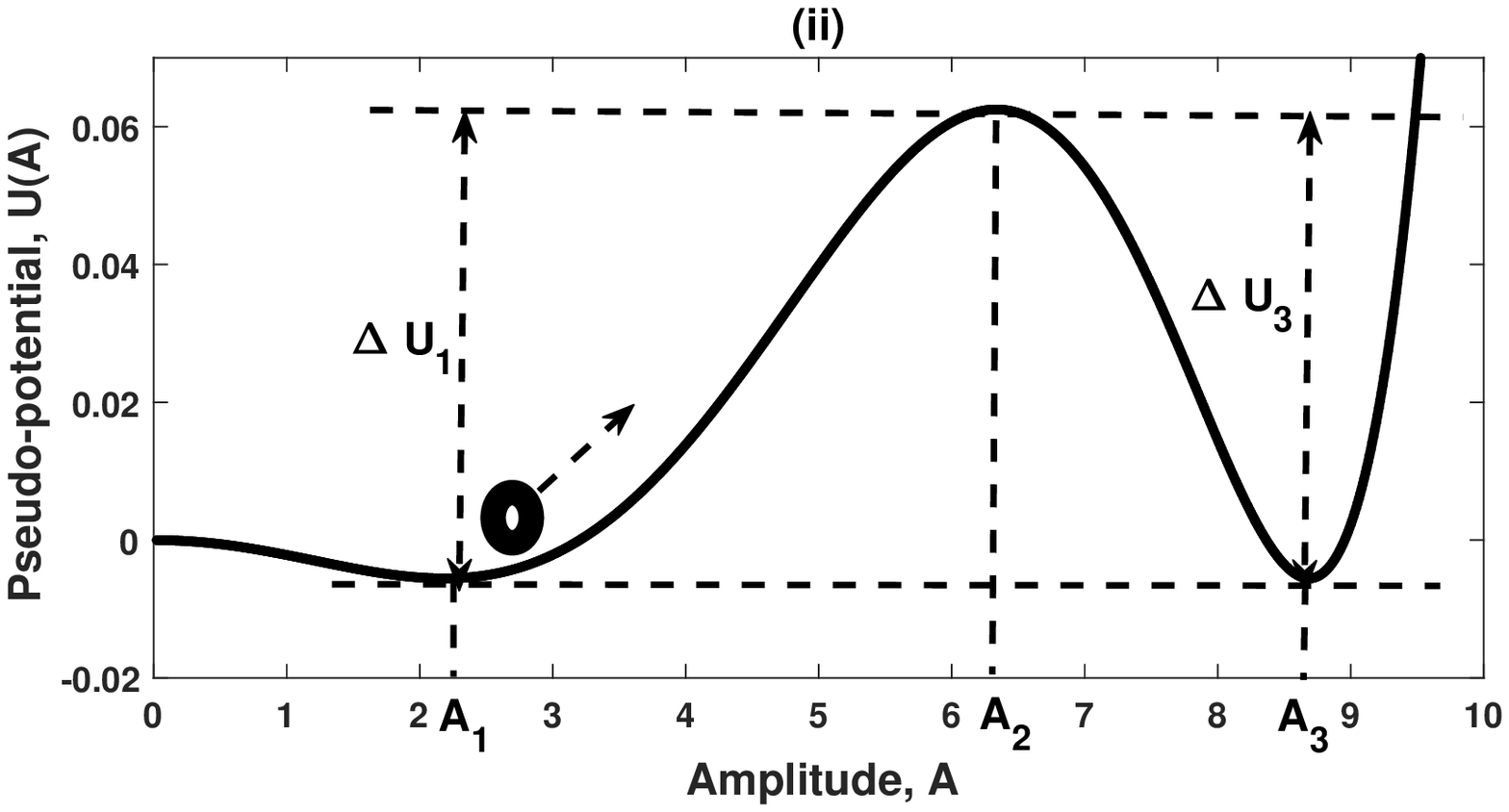}
\caption{\it The effective pseudo--potential  $U(A)$ versus the amplitude $A$
for the free-noise self-sustained oscillator with
multi-limit-cycles, while (i) correspond to the asymmetric potential with the $AS_1$ parameters
and (ii) to the symetric potential with the $SS_1$ parameters. The other parameter is $\mu=0.01$.
}
\label{fig2}
\end{center}
\end{figure}

The random variables $\xi$ corresponding to the characteristic
 functions (\ref{eq5}) can be generated by the algorithm presented in Refs. \cite{r4,r5} as follows:
first one generates the random variables $U$ uniformly distributed on $[-\frac{\pi}{2},+\frac{\pi}{2}]$ and the variable $W$ exponentially distributed with a unit mean; $U$ and $W$ being statistically independent.
The L\'evy distributed variable $X$ can be generated as follows:\\
For $a\neq 1$:
\begin{equation}\label{eqeq6}
  X=S_{a,b}\frac{\sin( a(U+B_{a,b}))}{(\cos(U))^{1/a}}\left( \frac{\cos(U-a(U+B_{a,b}))}{W}\right)^{\frac{1-a}{a}},
\end{equation}
where
\begin{eqnarray}
\label{eq7}
  \left\{\begin{array}{l}
    B_{a,b}= \frac{\arctan(b\tan(a\pi/2))}{a},\\
    S_{a,b}=\left(1+(b\tan(a\pi/2))^2\right)^{\frac{1}{2a}}.
  \end{array}\right.
\end{eqnarray}
For $a=1$:
\begin{equation}\label{eq8}
  X=\frac{2}{\pi}\left[\left( \frac{\pi}{2}+bU\right)\tan U-b\log \left( \frac{a\pi W\cos U}{2bU+a\pi}\right) \right].
\end{equation}
Finally, the abovementioned $\xi$ reads:
\begin{eqnarray}
\label{eq9}
  \xi =
  \left\{\begin{array}{ll}
    \sigma X+\delta, & a\neq 1, \\
    \sigma X+2b\sigma \frac{\log (\sigma)}{2} +\delta,& a=1.
  \end{array}\right.
\end{eqnarray}
The L\'evy noise is a formal time derivative of the generalized Wiener process.
For the time step of integration $\Delta t$, the increments
of the generalized Wiener process are distributed
according to the distribution $L(\xi,\sigma \Delta t^{1/a},\delta)$.
The L\'evy process can be retrieved with the transformation $\zeta(t)=\Delta t^{1/a}\xi$ \cite{r4,r5}.
L\'evy probability densities functions under different stability indexes and skewness parameters are presented in Fig.\ref{fig1a}, symmetric for $b = 0$; for $a < 1$, $L(\zeta,\sigma\Delta t^{1/a},\delta)$, left--skewed for $b < 0$ and right--skewed for $b > 0$.
For $a > 1$, $L(\zeta,\sigma\Delta t^{1/a},\delta)$ is right--skewed for $b < 0$ and left-skewed for $b > 0$.

\subsection{The numerical method to simulate the birhythmic van der Pol system driven by Levy noise}

Let us consider the multi-limit-cycle van del Pol-like oscillator
 to model coherent oscillations in
biological systems, such as an enzymatic substrate reaction with
ferroelectric behavior in brain waves models
(see
Ref.\cite{kaiser1,frohlich,enjieu-chabi-yamapi-woafo} for more details).
 In this case, one should include the electrical field applied to the excited enzymes,
which depends for example on the external chemical influences ({\it
i.e.}, the flow of enzyme molecules through the
transport phenomena). One can therefore assume that the
external chemical influence and the dielectric contain a random perturbation.
Therefore, adding both the chemical and the dielectric
contribution, the activated enzymes are subject to a random excitation
 governed by the Langevin version of Eq.~(\ref{eq1}), namely:
\begin{equation}
\label{eq10}
  \ddot x-\mu(1-x^2+\alpha x^4-\beta x^6)\dot x+x= \zeta(t),
\end{equation}
where $\zeta(t)$ denote L\'evy noise and is
the formal
time derivative of a L\'evy process $\xi (t)$, which can be viewed
as a generalized Wiener process, obeying to the L\'evy distribution $L(\xi,\sigma \Delta t^{1/a},\delta)$.
We recall that the L\'evy noise measure the intensity of the random electrical field.
By introducing
the new variable $\dot x = u$, equation (\ref{eq10}) can be write in the form:
\begin{eqnarray}
\label{eq11}
\left\{
  \begin{array}{l}
  \dot x=u,   \\
   \dot u=\mu(1-x^2+\alpha x^4-\beta x^6)u-x+\zeta(t),
  \end{array}
\right.
\end{eqnarray}
 and the relative difference scheme \cite{r1} is obtained to calculate:
system (\ref{eq11}):
\begin{eqnarray}
\label{eq12}
\left\{
  \begin{array}{l}
x_{n+1}-x_{n}=u_{n}\Delta t, \nonumber \\
u_{n+1}-u_{n}=[\mu(1-x_{n}^2+\alpha x_{n}^4-\beta x_{n}^6)u_{n}-x_{n}]\Delta t
+\Delta t^{1/a}\xi,
  \end{array}
\right.
\end{eqnarray}
 where $\xi$ denotes L\'evy distributed random number with the stability L\'evy
index $a$ and the noise intensity $D = \sigma^a$.
For the sake of simplicity and uniform, all our simulations are
performed with the time step $\Delta t = 0.01$.
In this work we only consider the case $\delta = b = 0$, \emph{i.e.} the symmetric L\'evy distribution.
%\section{Effective quasi-potential in the presence of Levy's noise}
%\subsection{Analytical estimation of the quasi-potential}
%\subsection{Influence of the Levy's noise parameters}

\section{Global stability analysis}

A principal question about the effects of noise is the occurrence
of large deviations, that is excursions from an attractor to another.
In fact, an attractor is only locally stable, but for birhythmicity to be actually displayed one
might be interested in the global analysis, that is the time spent on average in the proximity of each attractor.
The approach for potential systems, that is when the force can be derived from the gradient of a function, one can refer to the classical Kramers theory, with the many modifications that have been developed.
For non-gradient systems, the quasi-potential plays a similar role, for it determines the asymptotic low noise limit for Gaussian noise \cite{Graham85}.
The quasi-potential has proved effective for van der Pol birhythmic system driven by uncorrelated and correlated Gaussian noise, also in the presence of a sinusoidal forcing term \cite{Mbakob1,Mbakob2,Yamapi12,Yamapi17}.
Our goal is to determine whereas the same approach can be effective for L\'evy noise.

\subsection{Statement of the problem}
\label{global}

A quasi-potential function $U(A)$ is an effective energy in the sense that it
 determines in the low noise regime, the escapes from the attractor with an Arrhenius-like behavior  \cite{Yamapi10}:
\begin{equation}
\label{Arrhenius}
\langle T \rangle \propto \exp (\Delta U/D).
\end{equation}

A number of questions arise in the extension of Eq.(\ref{Arrhenius})
to birhythmic van der Pol under L\'evy noise influence.
First, one can ask how to modify the functional form of Eq.(\ref{Arrhenius}).
The escapes are governed, for L\'evy noise
systems with a bona fide potential \cite{Chechkin05,Chechkin07,Guarcello18}:
\begin{equation}
\langle T_{esc} \left ( a,D \right ) \rangle =
\left (\frac{ \eta ^{1-\mu_{a}} \Delta x^{2-2\mu_{a} + a \mu_{a}} }
 {4^{1-\mu_{a}} \Delta U^{1-\mu_{a}} 2^{a \mu_{a}}} \right )
\frac{{\cal C}_{a}}
{D^{\mu_{a} }},
\label{tau_Levy_gen}
\end{equation}
where $\Delta x$ is the distance between the stable minimum of the potential and the separatrix, $\Delta U$ is the energy or activation barrier, $\eta$ is the damping index (that only appears to correctly normalize the overdamped equation), $a$ the index of the L\'evy distribution, $D$ the noise intensity.
The scaling exponent $\mu_{a}$ and the coefficient ${\cal C}_a$ are supposed to have a universal behavior for overdamped  systems~\cite{Chechkin05}, such as:
\begin{equation}
\label{mutheoretical}
\mu_{a} \simeq 1+ 0.401 \left(a - 1\right) + 0.105 \left(a - 1\right)^2.
\end{equation}
The adaptation of Eq.(\ref{tau_Levy_gen}) to the van der
Pol birhythmic system gives a first analytical result on the influence of L\'evy noise.
In other words, in Eq.(\ref{tau_Levy_gen}) $x$ is a generic coordinate where the force stems from a potential.
As such, $x$ cannot be the variable of Eq.(\ref{eq5}),
that is a non-potential (or non-gradient) ordinary differential equation, and hence $U$ does not exist.
To apply the theory of L\'evy noise \cite{chechkin1,chechkin2} to Eq.(\ref{eq5}) it is necessary to introduce a quasi-potential that playes the role of $U$.
Also, the dynamic variable can presumably change, and is not any more $x$.

The form of the quasi-potential has been derived with the stochastic averaging method for Gaussian noise.
How to calculate the same potential for L\'evy noise is still an open problem.
As a first guess, let us assume that it is the same potential as in the case of the Gaussian noise:
\begin{eqnarray}
\label{eqpot}
\frac{dA}{dt} &=& -\frac{dU(A)}{dA} +\sqrt{\tilde{D}}\zeta_\mathrm{1}(t),
\end{eqnarray}
where $A$ is a generic coordinate, e.g. the amplitude $A$  of the oscillations
as per Eq.(\ref{eq2}), $\zeta_1(t)$ is the Gaussian noise, $\tilde{D} $
an effective noise amplitude ($\tilde{D} = D/{\omega^{2}}$
in Ref. \cite{Mbakob2}) and the effective potential $U(A)$ is given by \cite{Mbakob2}:
\begin{eqnarray}
\label{eqU}
  U(A)= \frac{\mu}{128}\left(\frac{5\beta}{8}A^8 - \frac{4\alpha}{3}A^6 + 4A^4 - 32A^{2}\right) - \frac{\tilde{D}}{2}\ln(A).
\end{eqnarray}
As mentioned, a first rough approach could be to assume $D=0$ in Eq.(\ref{eqU}).
This choice corresponds to take the deterministic averaging.
The first possibility is therefore the following: to use Eq.(\ref{tau_Levy_gen}),
where the amplitude $A$ replaces $x$, and consider $U(A)$ as the quasi-potential.
In this approximation, $\Delta x$, the distance with the separatrix,
becomes $A_2-A_1$ and $A_3-A_2$ for the outer and inner barriers, respectively.
In this approximation, the damping reads $\eta=1$ .
Shortly, one could apply the theory of Checkin to Eq.(\ref{eqpot})
with $\zeta_\mathrm{1}(t)$ a L\'evy noise.
This is a very rough approximation, but gives an analytical prediction to be compared with numerical data.

There are other possibilities.
One is to use the principle of minimum action \cite{Sun17,Wan18}.
These authors describe a numerical method to derive the quasi-potential for a non gradient system.
The physical idea, that has also been used for Josephson junctions \cite{Kautz88,Kautz94},
 is that noise activated trajectories have a different weight, and that
 the minimum energy is the most likely to be followed by the noise driven system.
However, this line of research is numerically heavy, and might
also prove not appropriated for L\'evy noise, where the distance rather than the energy barrier matters.

A more promising avenue is perhaps to repeat the calculations
for the ordinary van der Pol system (with L\'evy noise) of Ref. \cite{Hogele14}.
In particular in the paper ''The exit problem from the neighborhood of a global attractor for
heavy-tailed L\'evy diffusions'', in Sect. 2.4 it is presented the application of the method.
Interestingly, in a paper on a similar subject, ''Metastability of Morse-Smale
dynamical systems perturbed by heavy-tailed L\'evy type noise'', the
same authors also consider the birhythmic system of Moran and Goldbeter \cite{Moran84}, Sect. 2.4.
The calculations on this birhythmic system with two attractors
is specifically considered to derive a quasi-potential rather than an ordinary potential.

  Finally, one could use the numerical method already employed for Gaussian noise, that is to revert the logic of  Eq. (\ref{tau_Levy_gen}) to determine the energy activation.
This approach leads to the following definition of quasi-potential:
\begin{equation}
 \Delta U \equiv
\frac{\eta }{4}
\left (
\frac{  \Delta x^{2-2\mu_{a} + a \mu_{a}} }
 {\langle T_{esc} \left ( a,D \right ) \rangle   2^{a \mu_{a}}}
 \right )^{\frac{1}{1-\mu_{a}}}
\left(
\frac{{\cal C}_{a}}
{D^{\mu_{a} }}
\right)^{\frac{1}{ 1-\mu_{a}}}.
\label{tau_Levy_pot}
\end{equation}
This procedure is rather cumbersome, for L\'evy escapes are almost independent of the potential height, that only appears in the prefactor of Eq.(\ref{tau_Levy_gen}).
It is anyway interesting to verify if, and to which extent, the behavior  of  Eq. (\ref{tau_Levy_gen}) is reproduced by the numerically retrieved confining energy.
%More detailed calculations, presumably with less but more accurate points on the horizontal   axis can confirm that the quasi-potential has the same features of a potential for the L\'evy noise.
The prefactor is the most delicate point in the calculations \cite{Hogele14}.
To underline the effects, one can rewrite Eq. (\ref{tau_Levy_gen}) as follows
%(that is the underline model of Figs. \ref{fig3},\ref{fig4},\ref{fig5})
:
\begin{equation}
\log\left[ \langle T_{esc} \left ( a,D \right ) \rangle \right] =
\log\left[
\frac { \eta ^{1-\mu_{a}} }
{4^{1-\mu_{a}} 2^{a \mu_{a}}}
\right]
+\log \left[ \frac
{ \Delta A^{2-2\mu_{a} + a \mu_{a}} }
{ \Delta U^{1-\mu_{a}}  }
 \right]
+\log\left[ {\cal C}_{a} \right]
-  \mu_{a} \log\left[ D \right].
\label{tau_Levy_log}
\end{equation}
As $\mu_a \simeq 1$ when the barrier is changed the main contribution arises from $\Delta A$.
This also could be checked numerically, using Tables \ref{Tab1},\ref{Tab2} for the distance $\Delta A$, possibly completed with the barrier heights $\Delta U$.

However, the very fact that the escapes follow the functional form of a power law (rather than an exponential), as shown in the numerical calculations
% of Figs. \ref{fig3},\ref{fig4},\ref{fig5},
is {\it per se} of interest.
%More numerical simulations might be used to support the claim and perhaps can carry information about the exponent $\mu_a$, to confirm that is almost independent of the L\'evy index, $a$.

\subsection{Escape times from the periodic attractors}

\begin{table}
\begin{center}
\begin{tabular}{|c|l|l|l|l|}
  \hline
  $Levy\,\, index\,\,a$ & $T_{esc}(A_1\longrightarrow A_3)$ && $T_{esc}(A_3\longrightarrow A_1)$ &\\
  \hline
  0.1  &  755.96&12.58&	1465.57&36.435\\
0.25  & 972.61	&&1669.03&\\
0.5   & 1510.22	&&2018.27&\\
0.75	&2540.22	&&1199.91&\\
1.0    &4064.31	&296.5&1603.60&1783.9\\
1.25	&5884.04	&&1261.42&\\
1.5    &10938.01	&&1032.77&\\
1.75   &24587.66	&&780.55&\\
1.9    &73130.41	&&589.92&\\
1.99   &1024791.23	&1118979.9&482.99&39468200.705\\
2.0    &14650719.12	&&492.74&	\\
  \hline
\end{tabular}
 \caption{\it{Escape times for the asymmetric pseudo-potential with low noise intensity $D=0.001$.
 All data refer to the case $\mu=0.01$.}
  \\{\bf  GF: I assume these are the numerical escape times. Could you also insert the analytical estimates?}
  }
  \label{Tab3}
\end{center}
\end{table}

\begin{table}
\begin{center}
\begin{tabular}{|c|l|l|l|l|}
  \hline
  $Levy\,\, index\,\,a $& $T_{esc}(A_1\longrightarrow A_3)$ & &$T_{esc}(A_3\longrightarrow A_1)$ &\\
  \hline
  0.1&	772.78&16.70&	1635.57&23.39\\
0.25&	1002.05&&	2321.54&\\
0.5&	1737.14	&&7259.01&\\
0.75&	2980.91	&&10192.21&\\
1.0	& 5377.62	&189.18& 13095.21&333.99\\
1.25&	9136.22&&	13629.12&\\
1.5	&20332.91	& &49020.82&\\
1.75 &	64860.04 &	&43477.14&\\
1.9	&36221.74&	&404335.44&\\
1.99 & 2326787.54&	18133.63&2957929.24&560387.519\\
2.0	&1422.24 &	&233281.25&\\
  \hline
\end{tabular}
 \caption{\it{Escape times for the symmetric pseudo-potential with low noise intensity $D=0.001$.
 All data refer to the case $\mu=0.01$.}
 \\{\bf  GF: I assume these are the numerical escape times. Could you also insert the analytical estimates?}}
  \label{Tab4}
\end{center}
\end{table}

\begin{figure}%[htbp]
\begin{center}
\includegraphics[height=5cm,width=8cm]{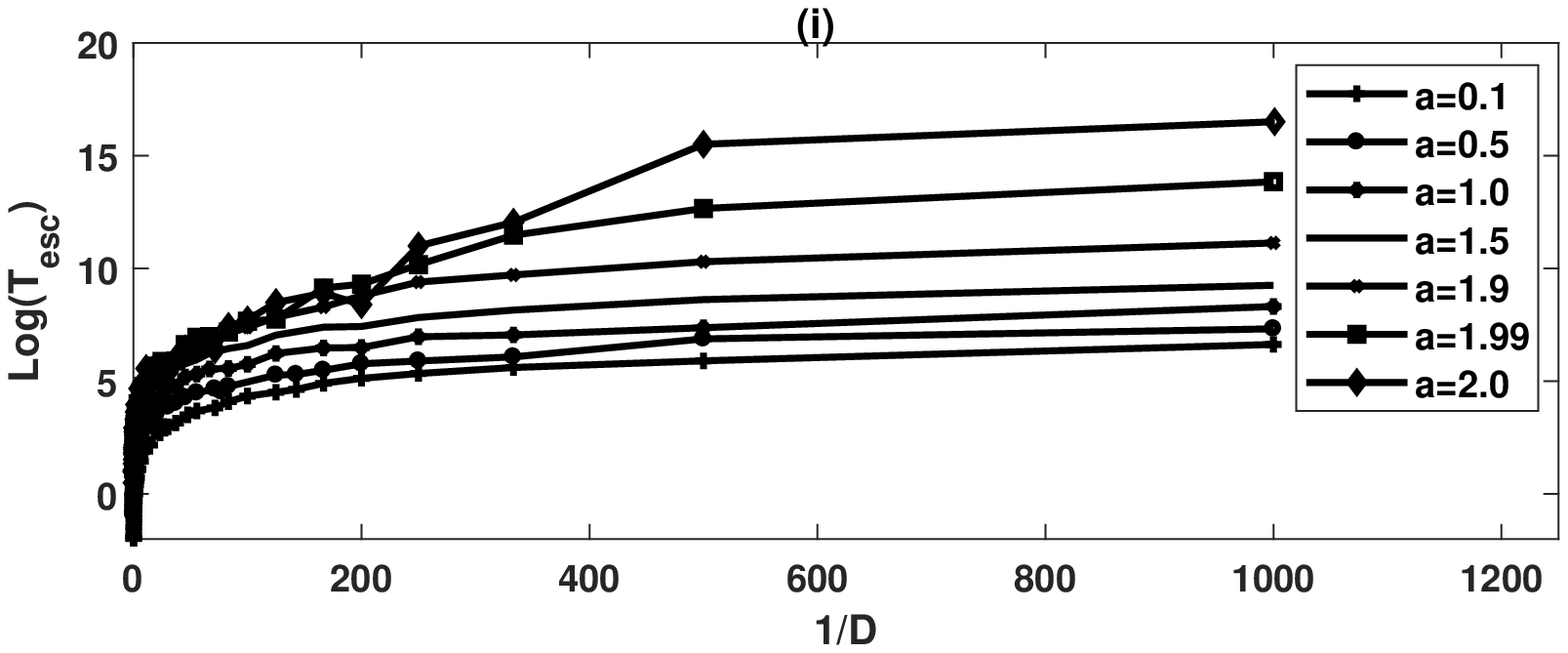}
\includegraphics[height=5cm,width=8cm]{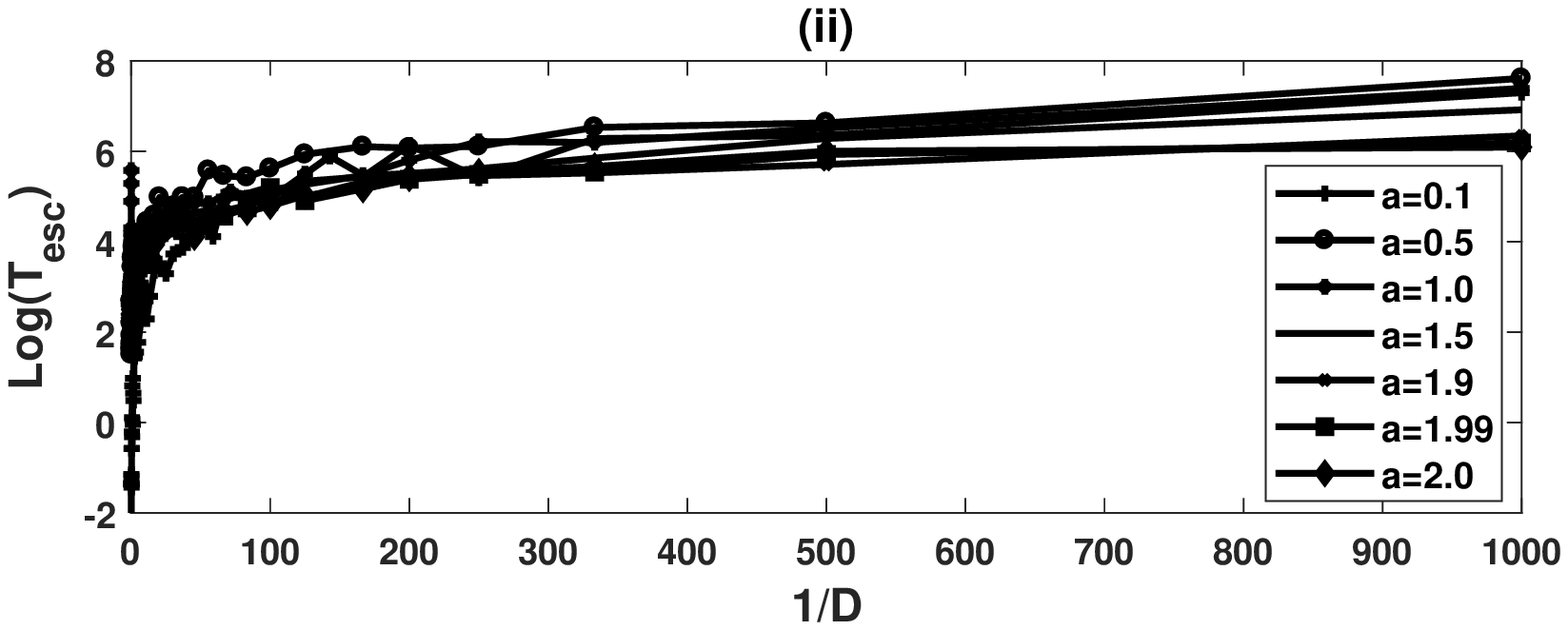}
\caption{\it Mean escape time versus the noise intensity $1/D$ for the asymmetric quasi-potential
with several different values of the L\'evy index, $a$,
(i) correspond of the escape transition $A_1\to A_3$ and (ii) for
the escape transition $A_3\to A_1$. The other parameter is $\mu=0.01$.
}
\label{fig3a}
\end{center}
\end{figure}

\begin{figure}%[htbp]
\begin{center}
\includegraphics[height=5cm,width=8cm]{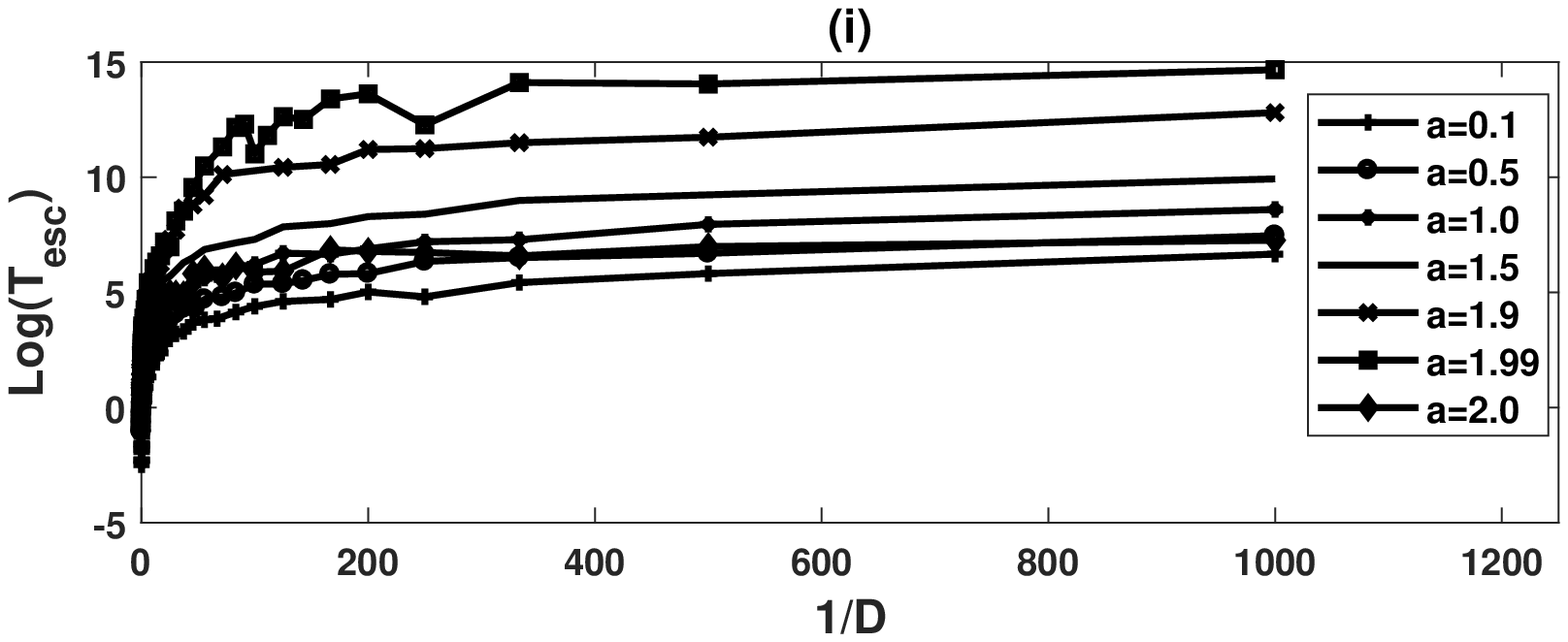}
\includegraphics[height=5cm,width=8cm]{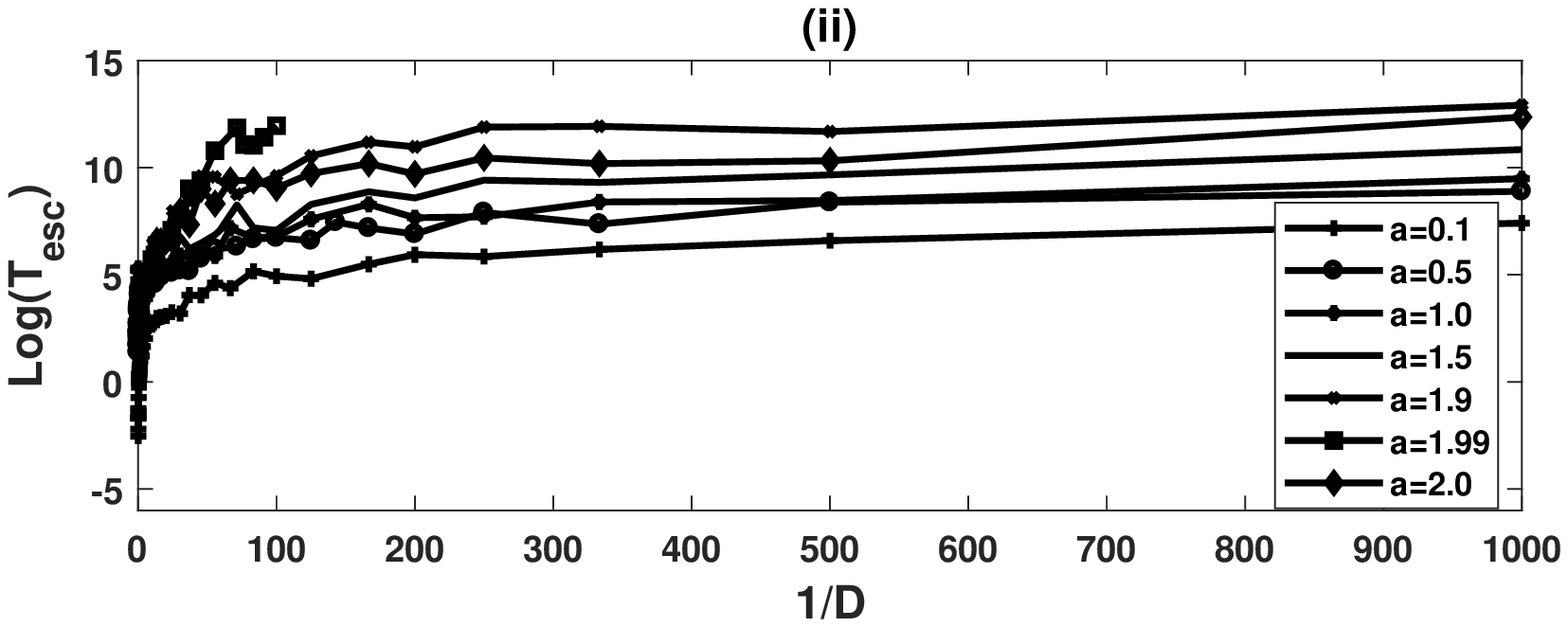}
\caption{\it
Mean escape time versus the noise intensity $1/D$ for the symmetric quasi-potential
with several different values of the L\'evy index, $a$,
(i) correspond of the escape transition $A_1\to A_3$ and (ii) for
the escape transition $A_3\to A_1$. The other parameter is $\mu=0.01$.}
\label{fig4a}
\end{center}
\end{figure}

\begin{figure}%[htbp]
\begin{center}
\includegraphics[height=5cm,width=8cm]{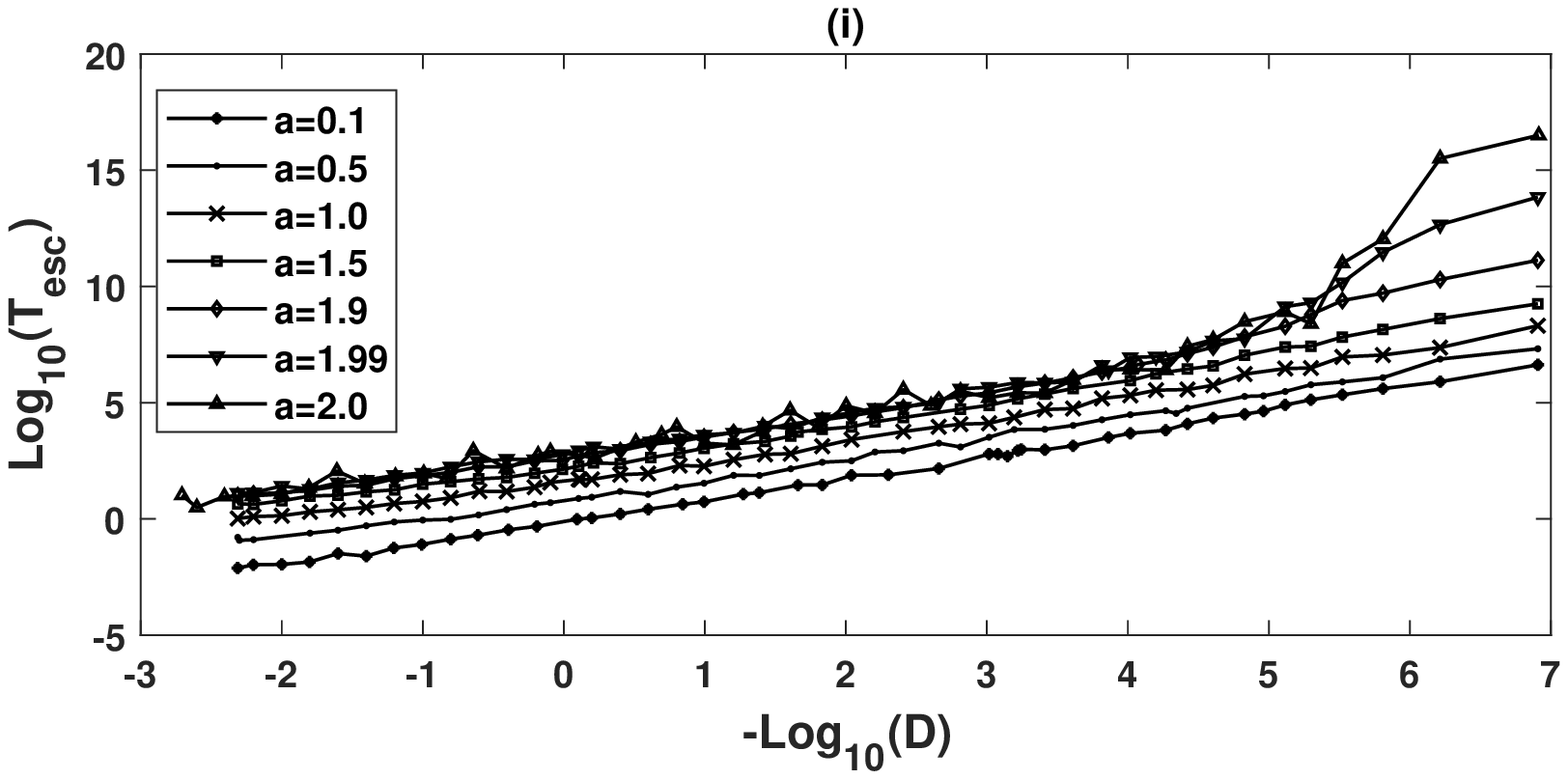}
\includegraphics[height=5cm,width=8cm]{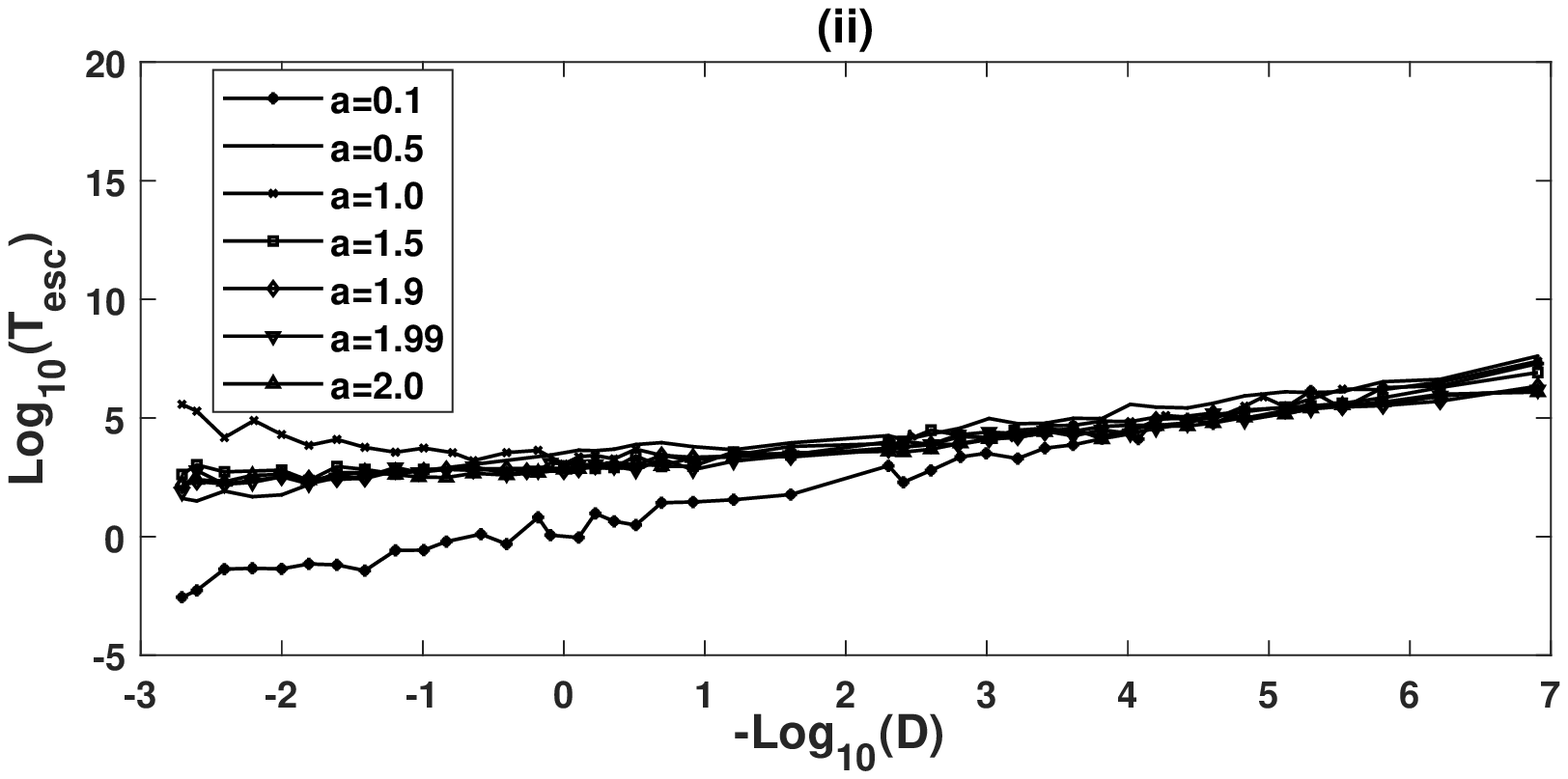}
\caption{\it Mean escape time versus the noise intensity $D$ for the asymmetric quasi-potential
with several different values of the L\'evy index, $a$,
(i) correspond of the escape transition $A_1\to A_3$ and (ii) for
the escape transition $A_3\to A_1$. The other parameter is $\mu=0.01$.
}
\label{fig3}
\end{center}
\end{figure}

\begin{figure}%[htbp]
\begin{center}
\includegraphics[height=5cm,width=8cm]{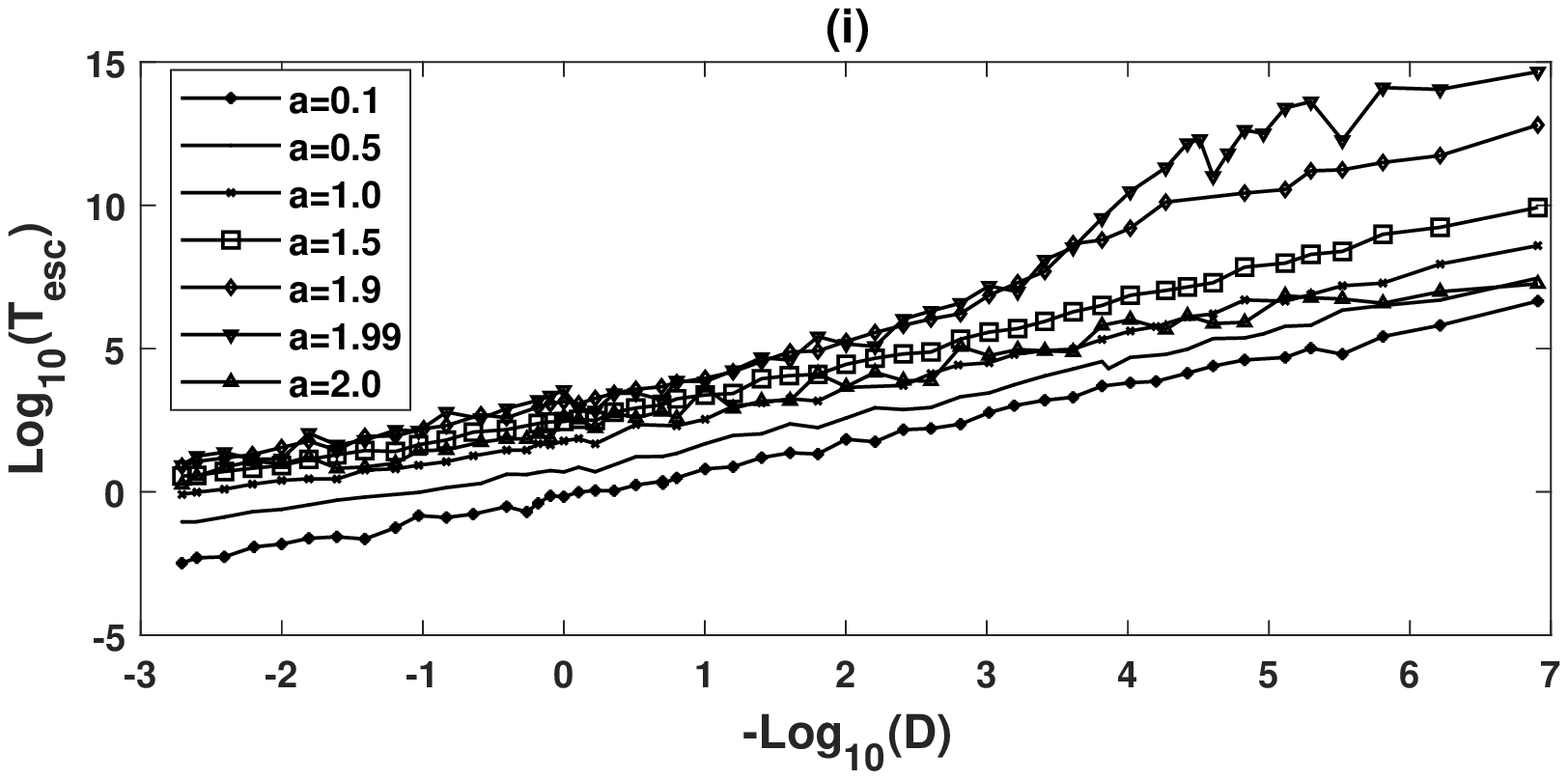}
\includegraphics[height=5cm,width=8cm]{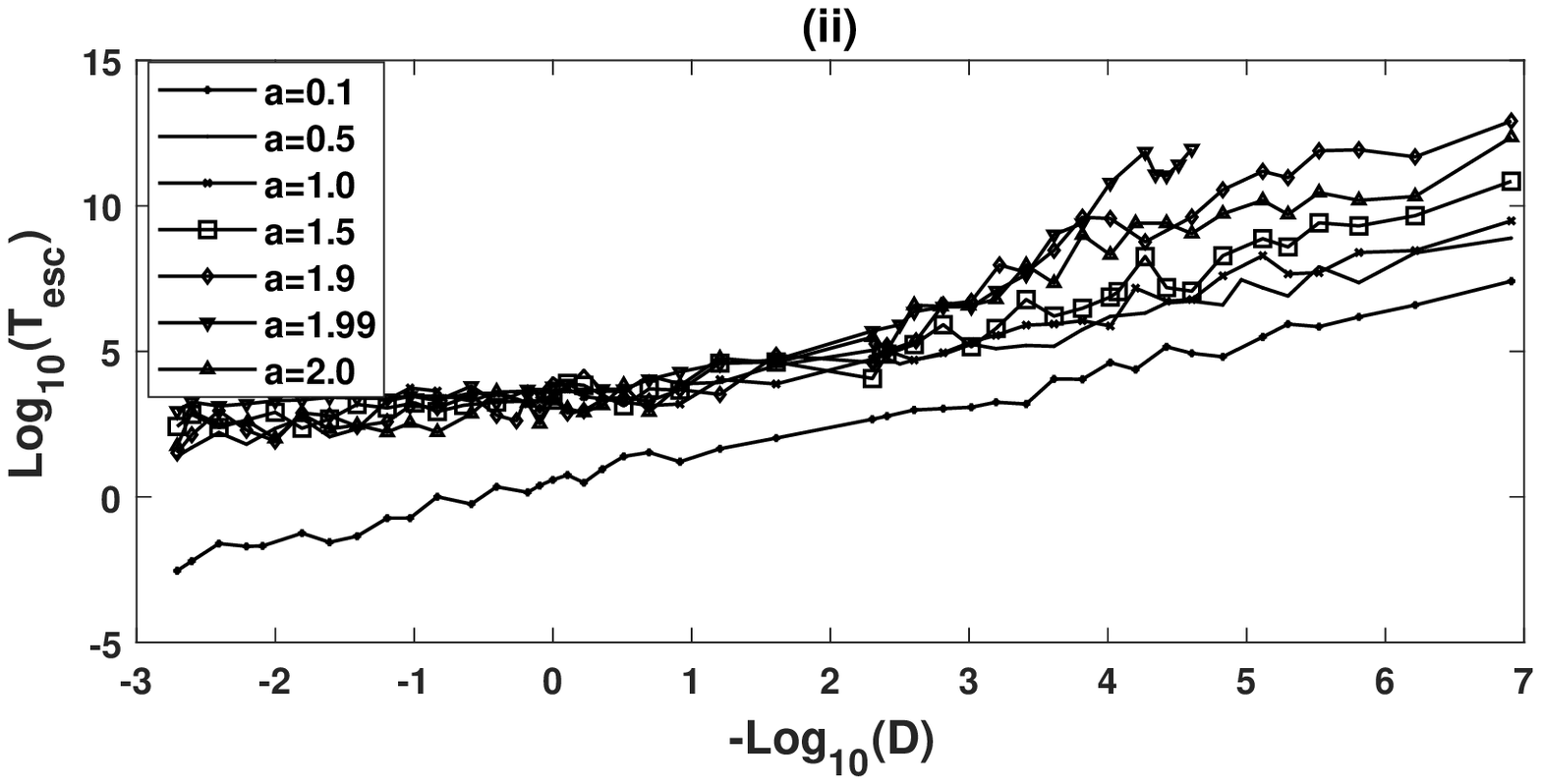}
\caption{\it
Mean escape time versus the noise intensity $D$ for the symmetric quasi-potential
with several different values of the L\'evy index, $a$,
(i) correspond of the escape transition $A_1\to A_3$ and (ii) for
the escape transition $A_3\to A_1$. The other parameter is $\mu=0.01$.}
\label{fig4}
\end{center}
\end{figure}

\begin{figure}%[htbp]
\begin{center}
\includegraphics[height=5cm,width=8cm]{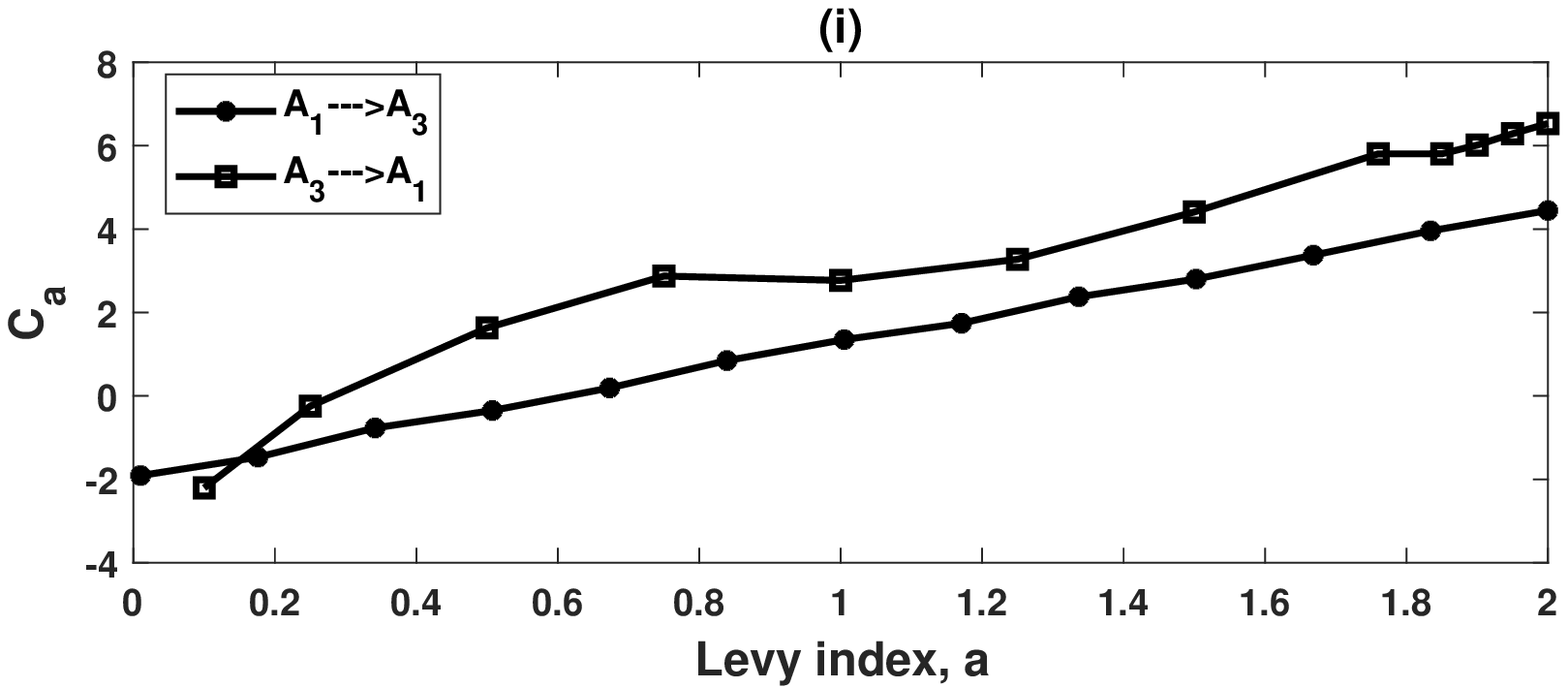}
\includegraphics[height=5cm,width=8cm]{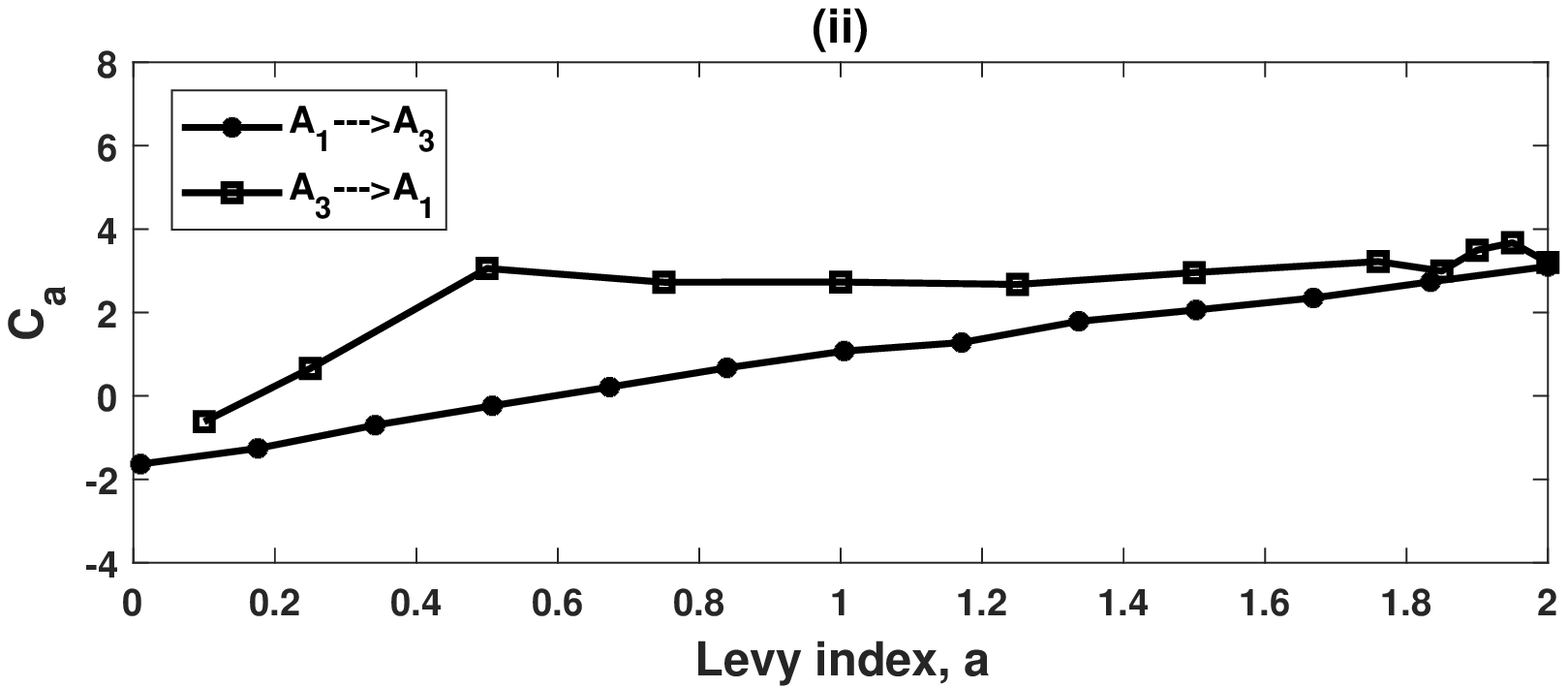}
\caption{\it Dependence of the scaling exponent $C_a$ versus the L\'evy index, $a$.
(i) correspond of the asymetric pseudo-potential and
 (ii) for the symetric pseudo-potential. The other parameter is $\mu=0.01$.}
\label{fig5}
\end{center}
\end{figure}

\begin{figure}%[htbp]
\begin{center}
\includegraphics[height=5cm,width=8cm]{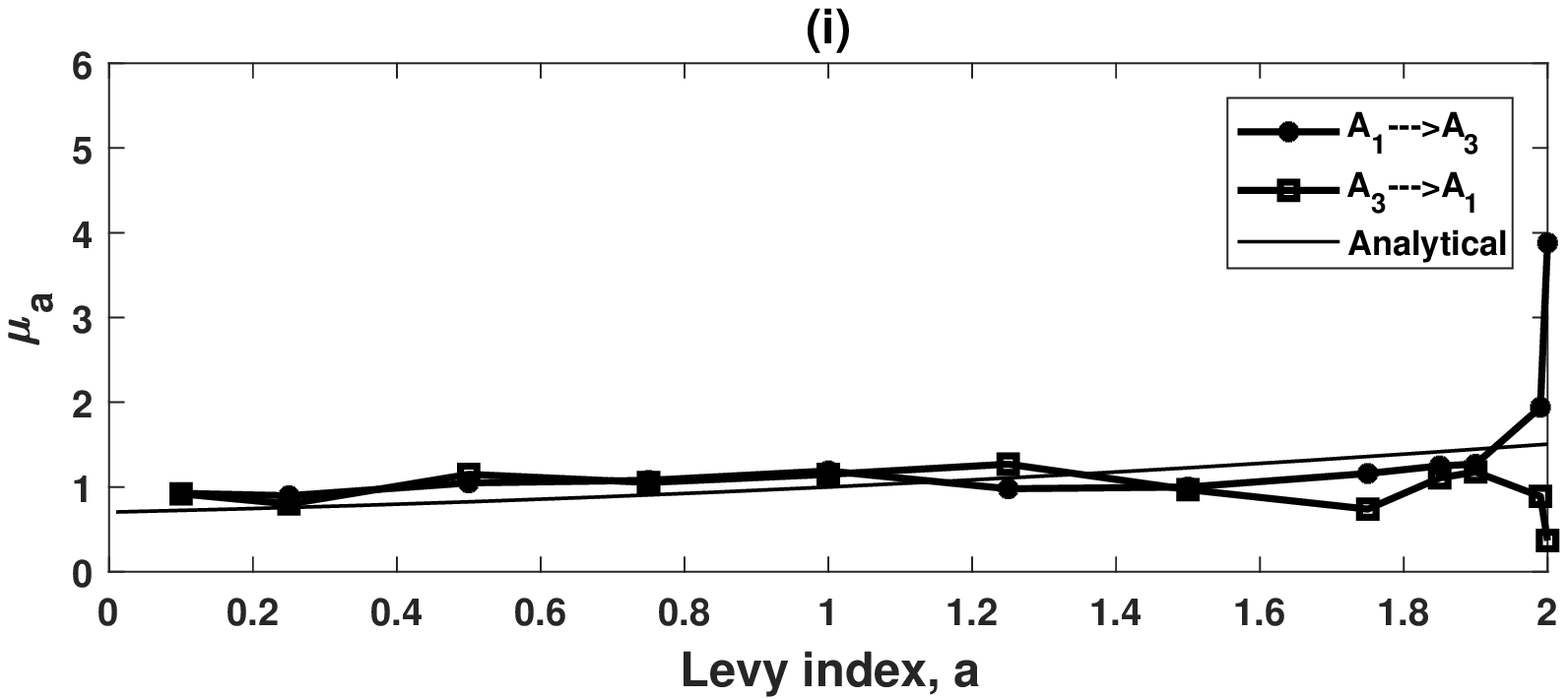}
\includegraphics[height=5cm,width=8cm]{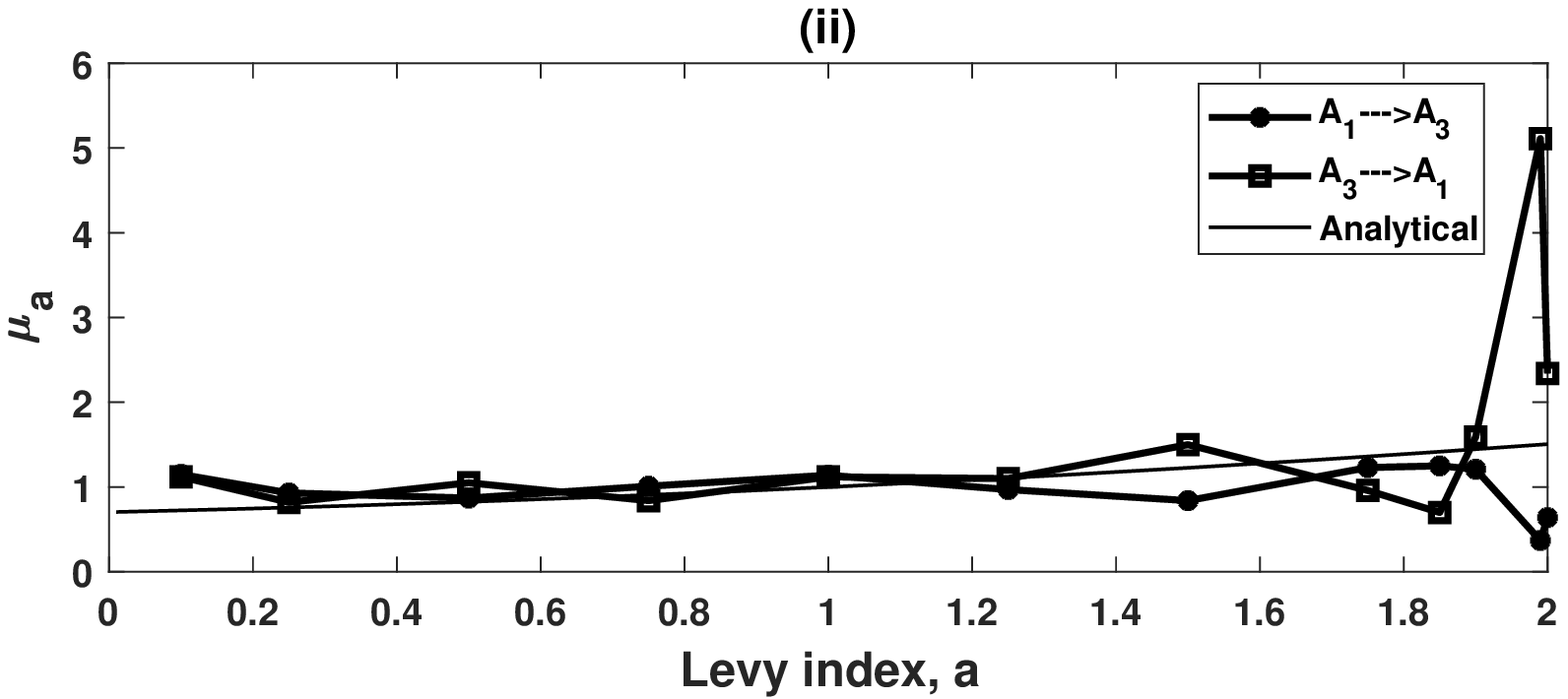}
\caption{\it Dependence of the scaling exponent $\mu_a$ versus the L\'evy index, $a$.
(i) correspond of the asymetric pseudo-potential and (ii) for the symetric pseudo-potential.
The analytic prediction of the solid line refers to Eq.(\ref{mutheoretical}).
\\{\bf GF: check the above statement is correct.}\\
The other parameter is $\mu=0.01$.}
\label{fig6}
\end{center}
\end{figure}

To examine the escape times from the periodic attractors caused by the L\'evy noise term in Eq.(\ref{eq9}) that induces the system to occasionally jump from one limit cycle to the other.
The system, initially on a limit-cycle attractor with amplitude $A_1$ or $A_3$, is forced by the random fluctuations  to leave the attractor and to wander about in the neighboring state space.
Escape occurs when this random motion drives the system across the boundary of the basin of attraction or  the unstable limit cycle with amplitude $A_2$, (\emph{i.e}.
$|x| > A_2$ (respectively, $|x| < A_2$)) over the activation energy  $\Delta U_1$ (respectively, $\Delta U_3$).
This energy is provided by the random force, that thus furnishes the energy equivalent to the depth of the left (respectively, right) well of the bistable potential.
The mean escape times $T_{esc}$ for the transitions $A_1\to A_3$ and $A_3\to A_1$ as function of the noise intensity $D$ for the asymmetric  and symmetric pseudo-potentials are shown in Figs. \ref{fig3} and \ref{fig4} for several different values of the L\'evy index $a$ ($0.1\leq a\leq 2$).
We generally observe that for both types of potential, the curves for $\alpha < 2$ obey
a different law in comparison to their Gaussian counterpart.
Earlier studies have shown that the variation of the escape time  will depend on the regime of the noise intensity to be considered.
 For  low noise intensity regime, in addition to the exponential dependence of the inverse of the intensity of the noise, $1 / D$, as it appears on Figs.\ref{fig3a},\ref{fig4a}.
We present in Figs. \ref{fig3},\ref{fig4}  a power-law  asymptotic behavior, as predicted by Eq. (\ref{tau_Levy_gen}).
 The figures refer to both the symmetric and asymmetric cases of the pseudo-potentials.
A thorough analysis of the data of the results presented in Figs. \ref{fig3} and
\ref{fig4} allows us to show the dependencies of the coefficient $C_a$
and the power-law exponent $\mu_a$ on the L\'evy index $a$in Figs.\ref{fig5} and \ref{fig6}, respectively.
In Fig. \ref{fig5} the coefficient $C_a$ theoretical behavior reads \cite{chechkin2} $1$
for $a\rightarrow 0$, passes through $\pi/2$ for $a=1$, and diverges as $1/(2-a)$
as the L\'evy parameter approaches $2$.
This qualitative behavior is very roughly reproduced by the data of Fig. \ref{fig5},
that does not start from $0$ as expected for $a \rightarrow 0$, reads $1$
instead of $4\pi/5$ for $a=1$, and diverges more mildly than expected.
Note that for our numerical investigation, $C_a$ reads $0.15$ for the index parameter
approaches $0.01$.
%{\bf  GF: check if you agree with my estimates, or if you can add a better comparison.
% Perhaps you can add points for the theoretical perdictions at $a=0$ and $a=1$.}
 In Fig.\ref{fig6}, the dependence of the scaling exponent $\mu_a$ versus the L\'evy index, $a$ are compared to the analytical approximation (\ref{mutheoretical}).
 For both symmetric and asymmetric pseudo-potential, the estimate is acceptable.
One can conclude that the prefactor $C_a$ is only qualitatively captured by the estimate for the ordinary potential, while the scaling exponent $\mu_a$ seems to be closer to the ordinary potential predictions.

\begin{figure}%[htbp]
\begin{center}
\includegraphics[height=5cm,width=8cm]{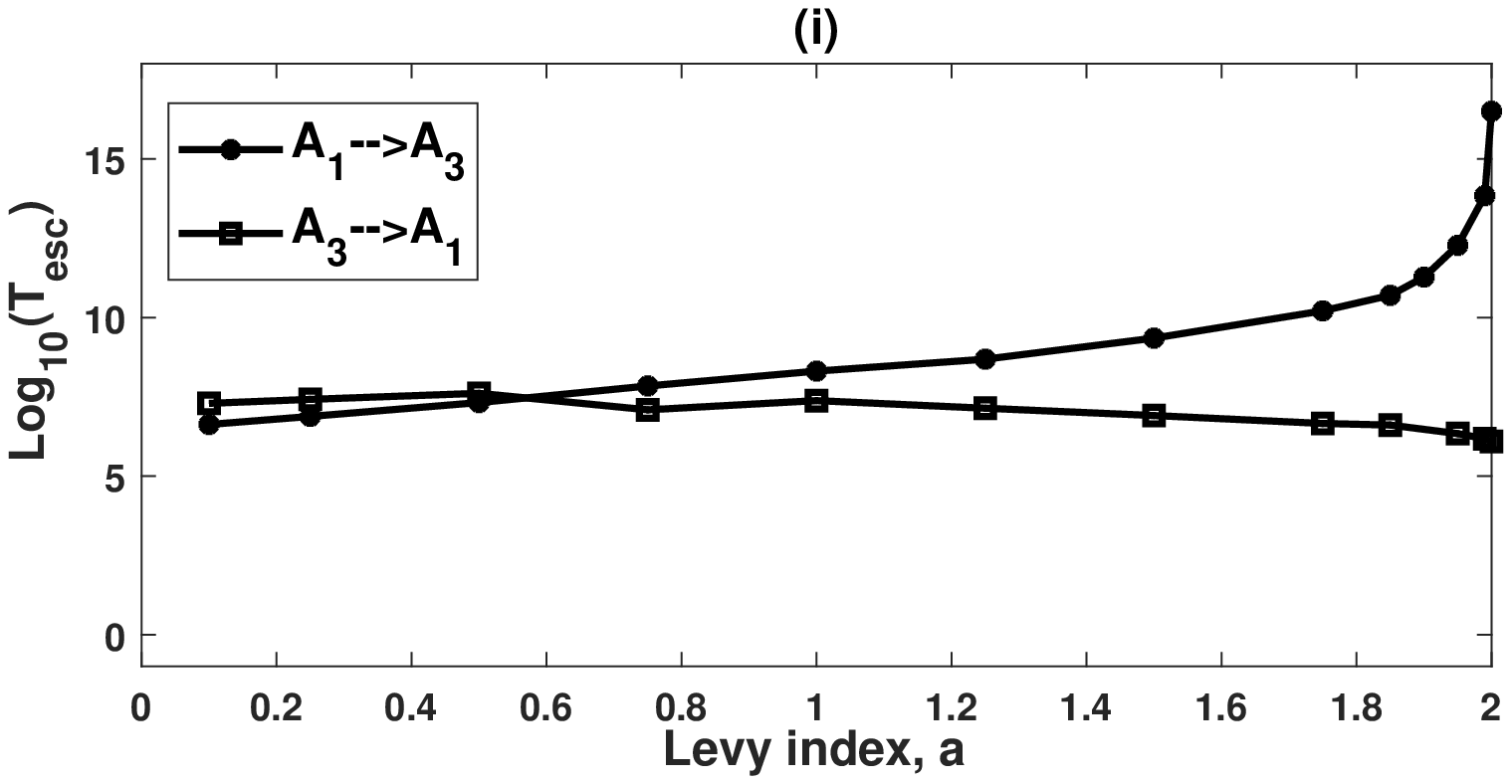}
\includegraphics[height=5cm,width=8cm]{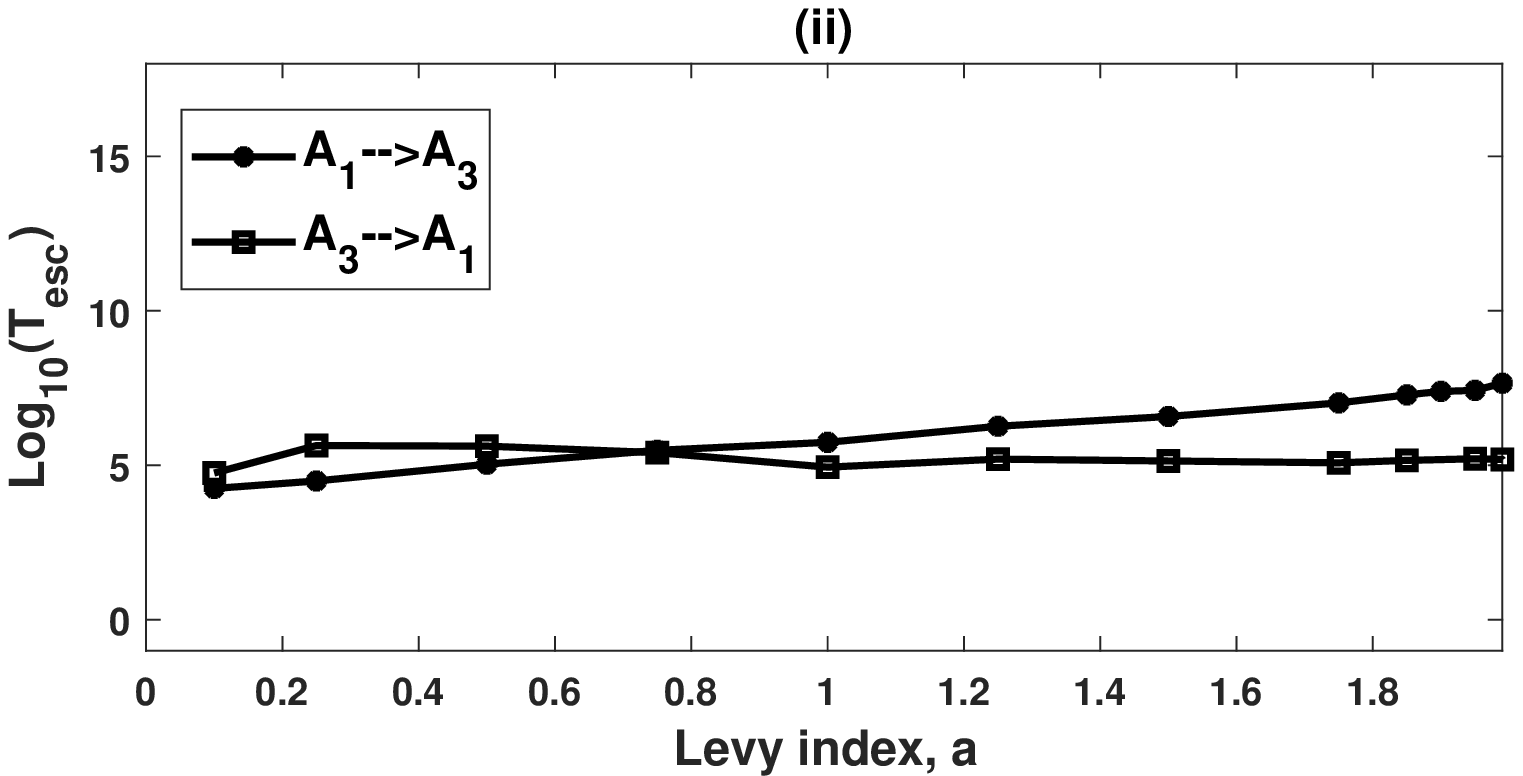}
\includegraphics[height=5cm,width=8cm]{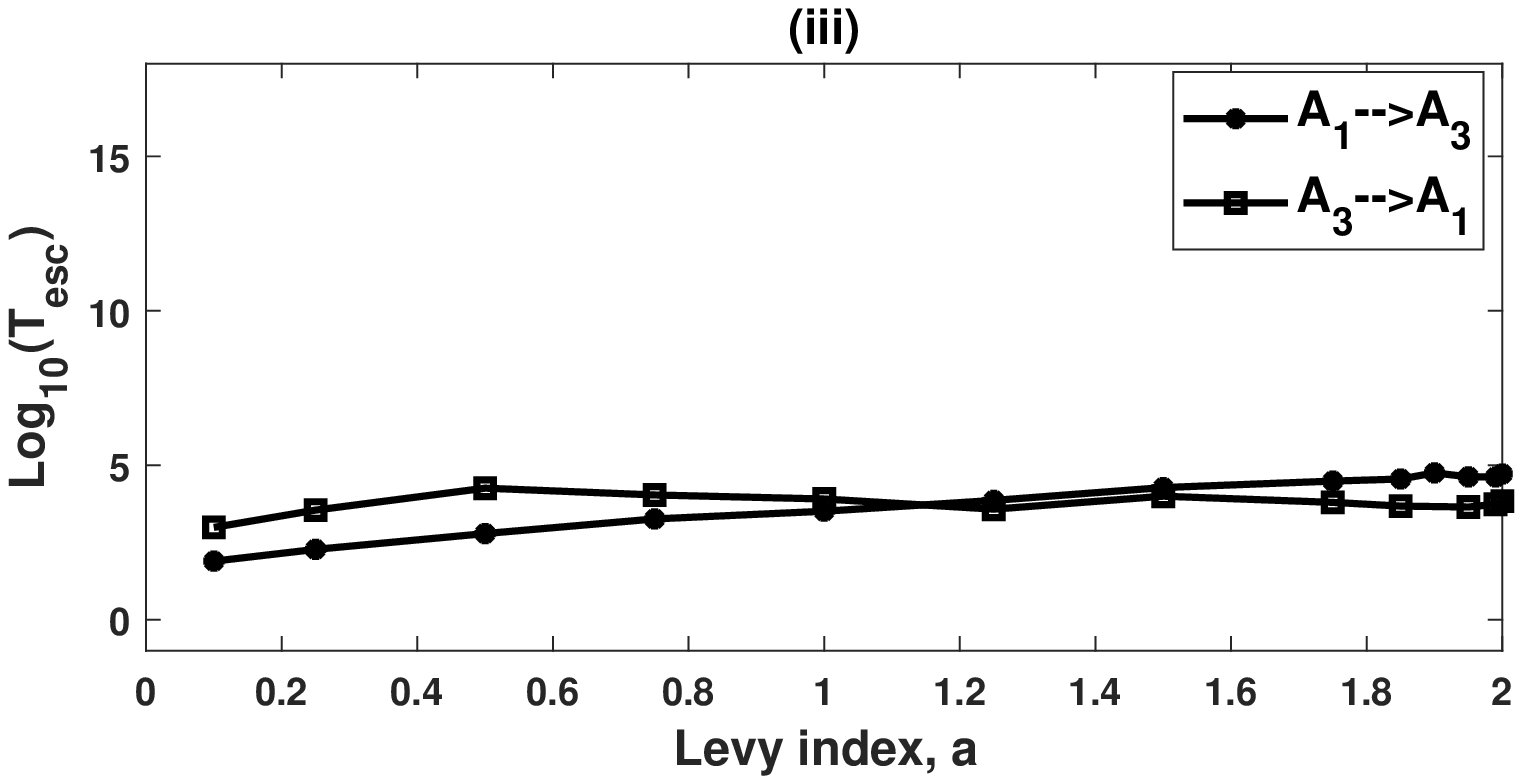}
\includegraphics[height=5cm,width=8cm]{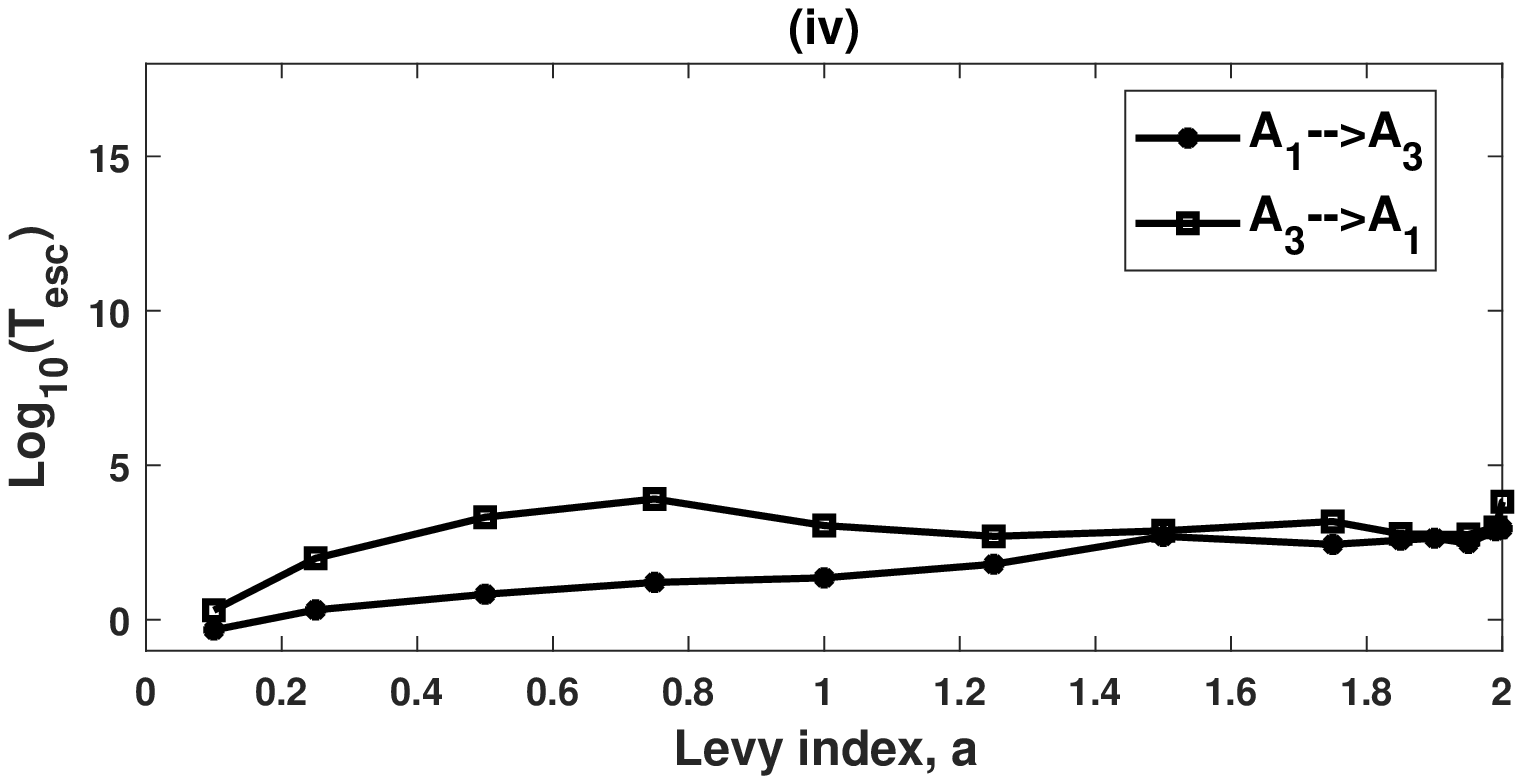}
\caption{\it Effects of the L\'evy noise intensity on the variation of the
 mean escape time versus the L\'evy index $a$ for the asymmetric quasi-potential (i):$D=0.001$;(ii):$D=0.01$; (iii):$D=0.1$;
(iv):$D=1.0$. The other parameter is $\mu=0.01$.
}
\label{fig7}
\end{center}
\end{figure}

\begin{figure}%[htbp]
\begin{center}
\includegraphics[height=5cm,width=7.8cm]{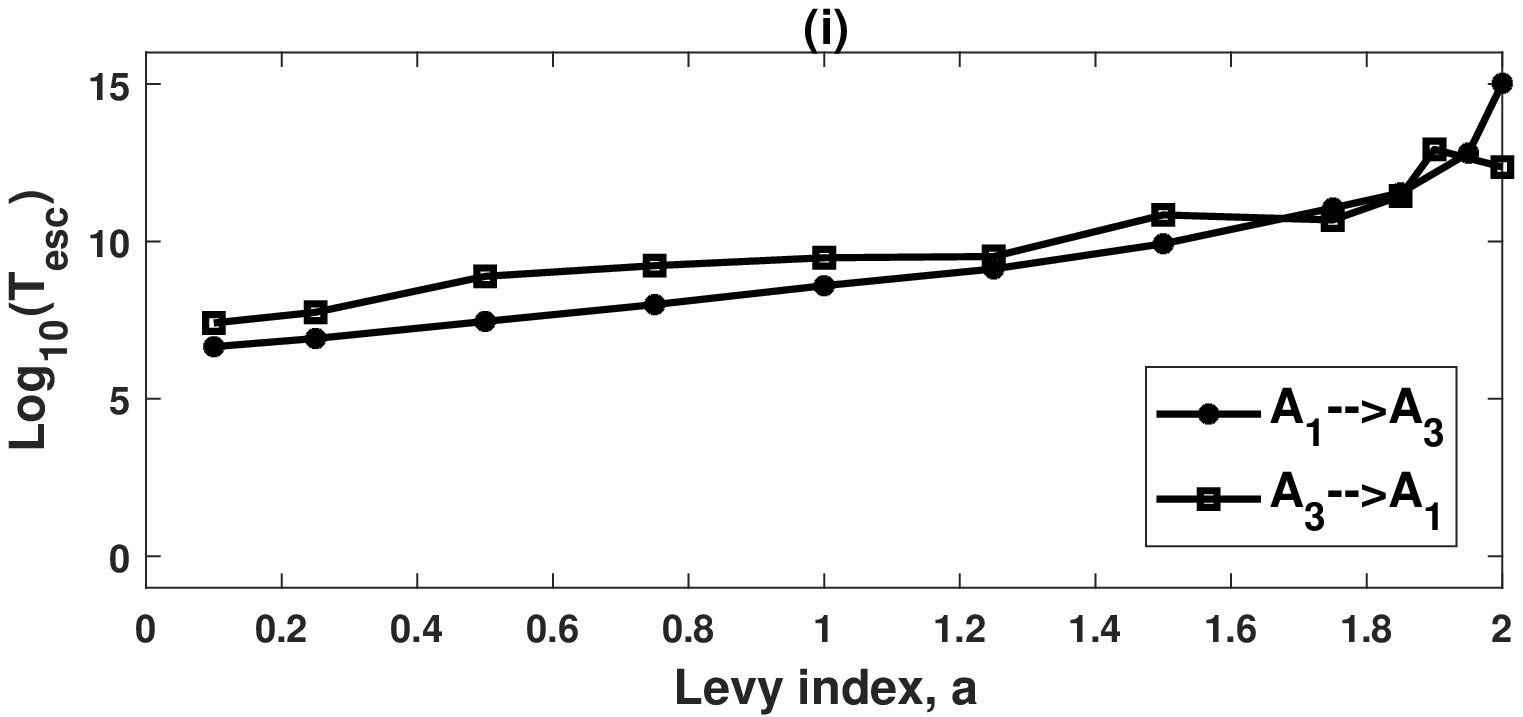}
\includegraphics[height=5cm,width=7.8cm]{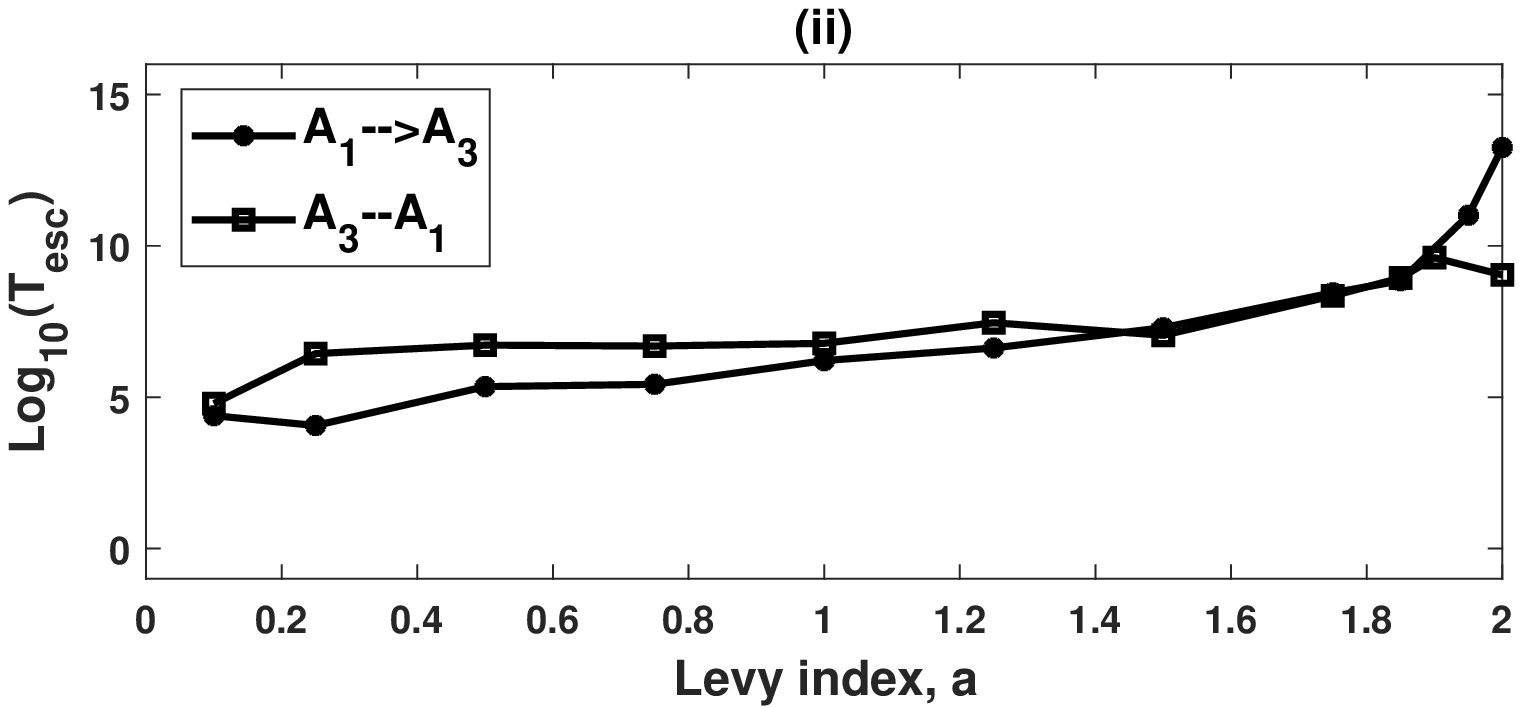}
\includegraphics[height=5cm,width=7.8cm]{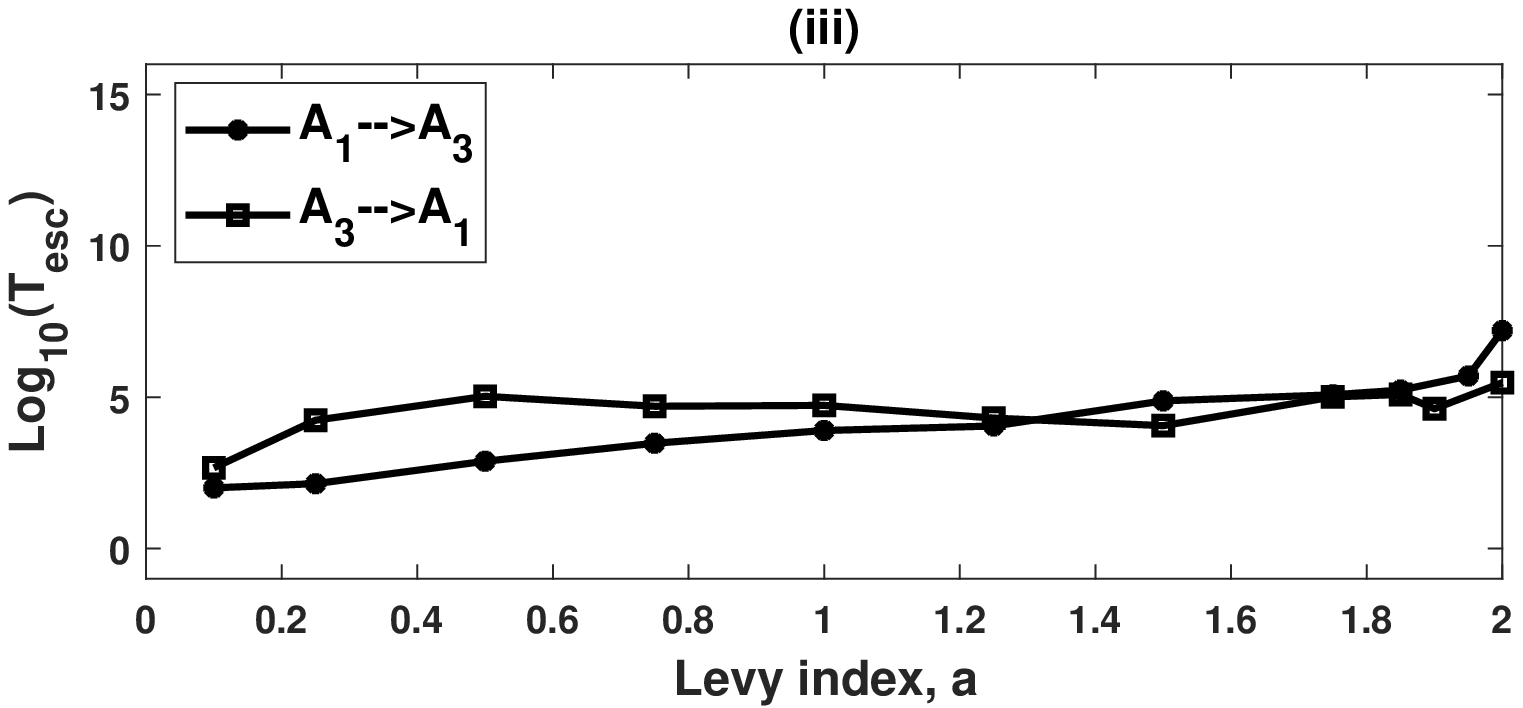}
\includegraphics[height=5cm,width=7.8cm]{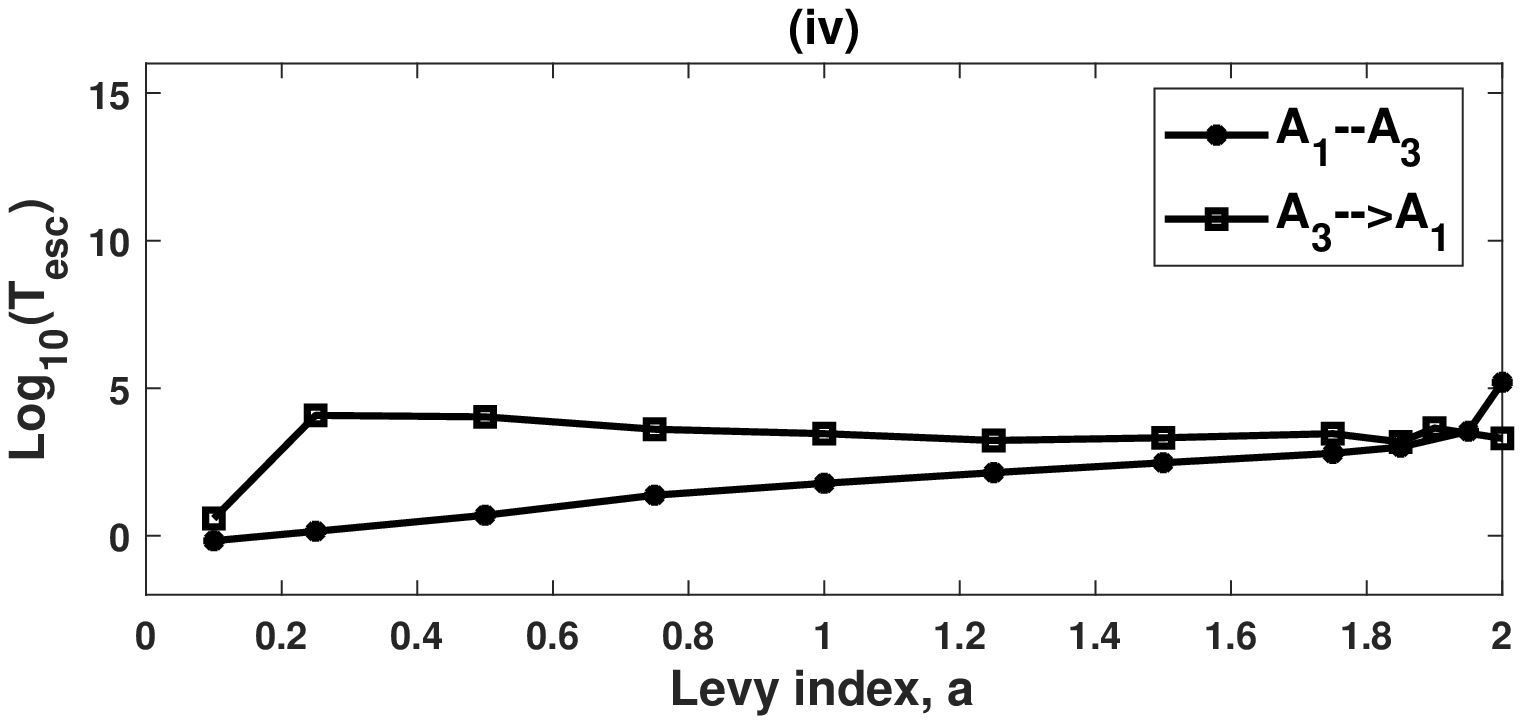}
\caption{\it Effects of the L\'evy noise intensity on the varaition of the
mean escape time versus the L\'evy index $a$ for the symmetric quasi-potential (i):$D=0.001$;(ii):$D=0.01$; (iii):$D=0.1$;
(iv):$D=1.0$. The other parameter is $\mu=0.01$.
}
\label{fig8}
\end{center}
\end{figure}

The influence of the noise intensity on the escape process is shown in Figs. \ref{fig7},\ref{fig8} as the dependence of the mean escape time $T_{esc}$ versus
the L\'evy index $a$, for several different values of the noise
intensity, $D$,  for the asymmetric and symmetric  pseudo-potential.

{\bf GF: could you add in the figures the predictions of Eq.(\ref{tau_Levy_log})?}

 As it is often the case, increasing the noise intensity reduces  the escape time in both types of potential.
 However, these observations  strongly depend on the L\'evy index $a$; it appears hat for a very small value of the intensity of the noise, for example $D = 0.001$,   the escape time increases or decreases considerably when the L\'evy index is close to $2$, that is close to the Gaussian case.
 For example, in the case of the asymmetric pseudo-potential, the escape transition time
  $T_{esc}(A_1\to A_3)$ (or $T_{esc}(A_3\to A_1)$ when the L\'evy index changes from
  $0.1$ to $2$, the escape time passes from $755$ to $14650719$ (from $1465$ to $492.4$, respectively).
This is reported in Table \ref{Tab3}, that displays the dependence   of the escape times as a function of the L\'evy index. In the case of symmetric pseudo-potential,
  one observes the same behavior when the L\'evy index increases, see Table \ref{Tab4}
  for a much more in-depth look at the dependence of escape time versus the L\'evy index, $a$.
It should be noted that in this case the pseudo-potential is initially symmetric without the L\'evy noise term, (\emph{i.e.} $T_{esc} (A_1 \to A_3)
\equiv T_{esc} (A_3 \to A_1)$), becomes asymmetric with the presence of this noise. It can be seen that when the L\'evy index $a$ increases from $0.1$ to $2.0$, the escape time to leave the pseudo-potential well around the limit cycle amplitude $A_3$ increases very considerably and makes the attractor $A_3$ more stable under the effect of the L\'evy noise.
As a result, a  particle confined in this pseudo-potential remains for a very long time under the effects of the random fluctuations induced by the L\'evy noise.

\begin{figure}%[htbp]
\begin{center}
\includegraphics[height=5cm,width=8.0cm]{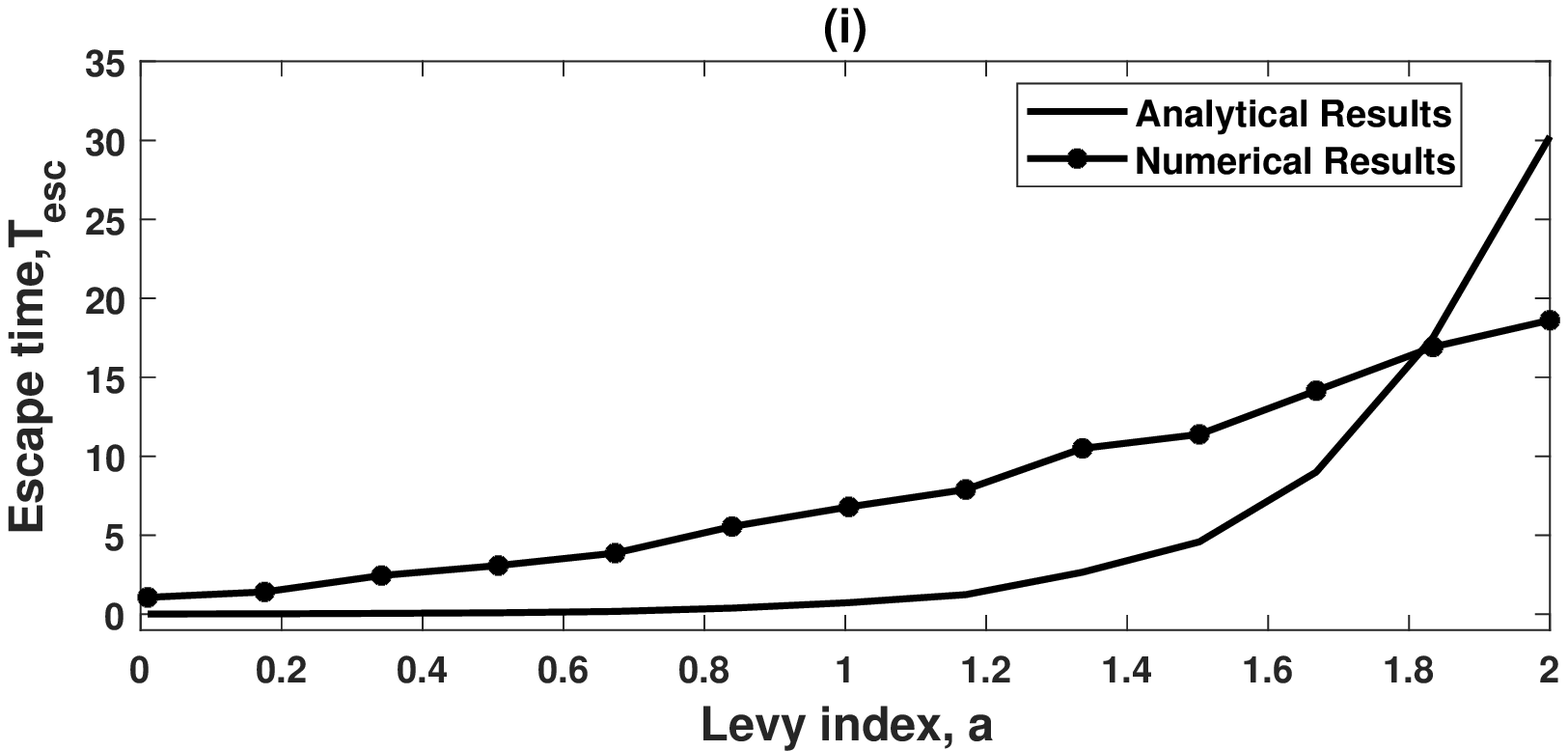}
\includegraphics[height=5cm,width=8.0cm]{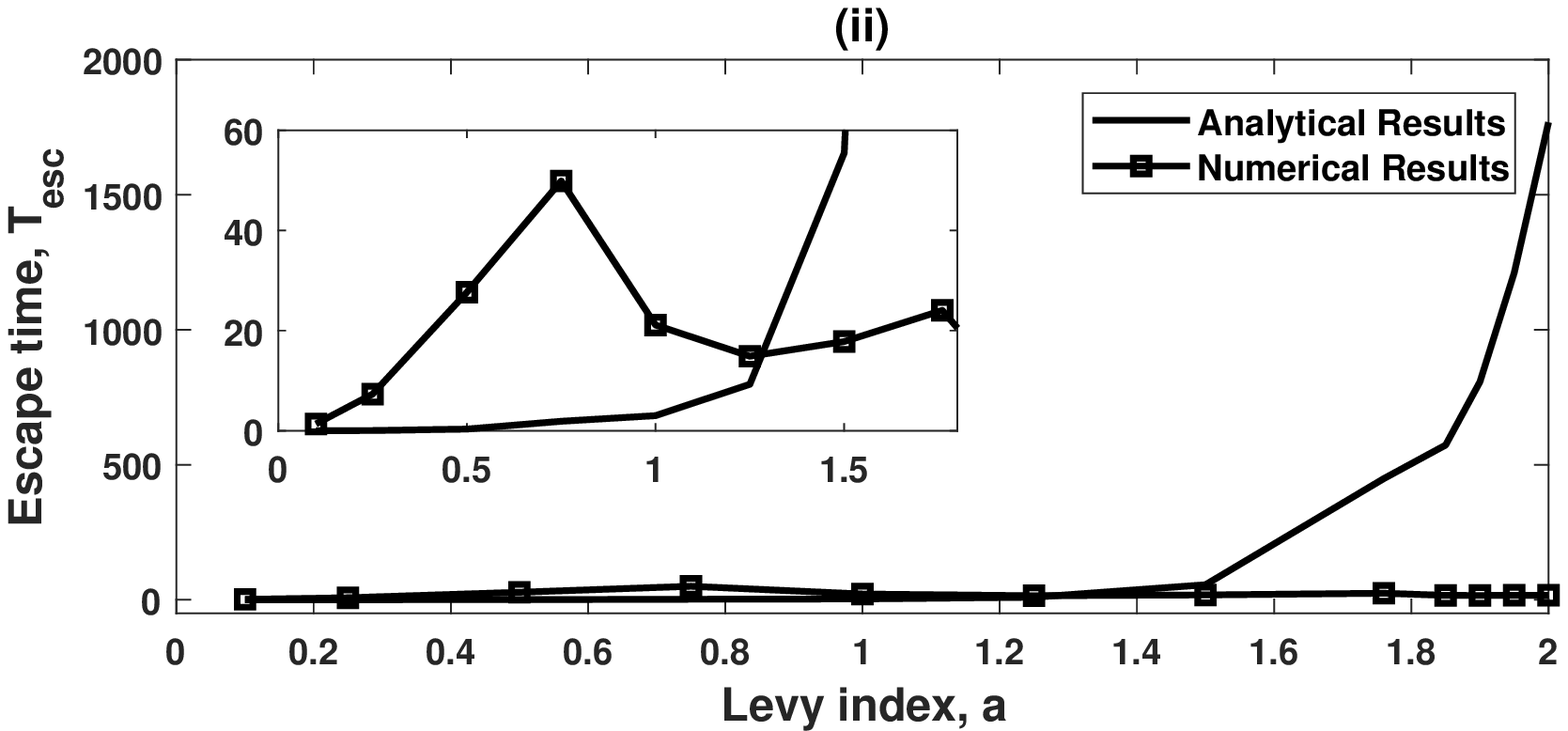}
\includegraphics[height=5cm,width=8.0cm]{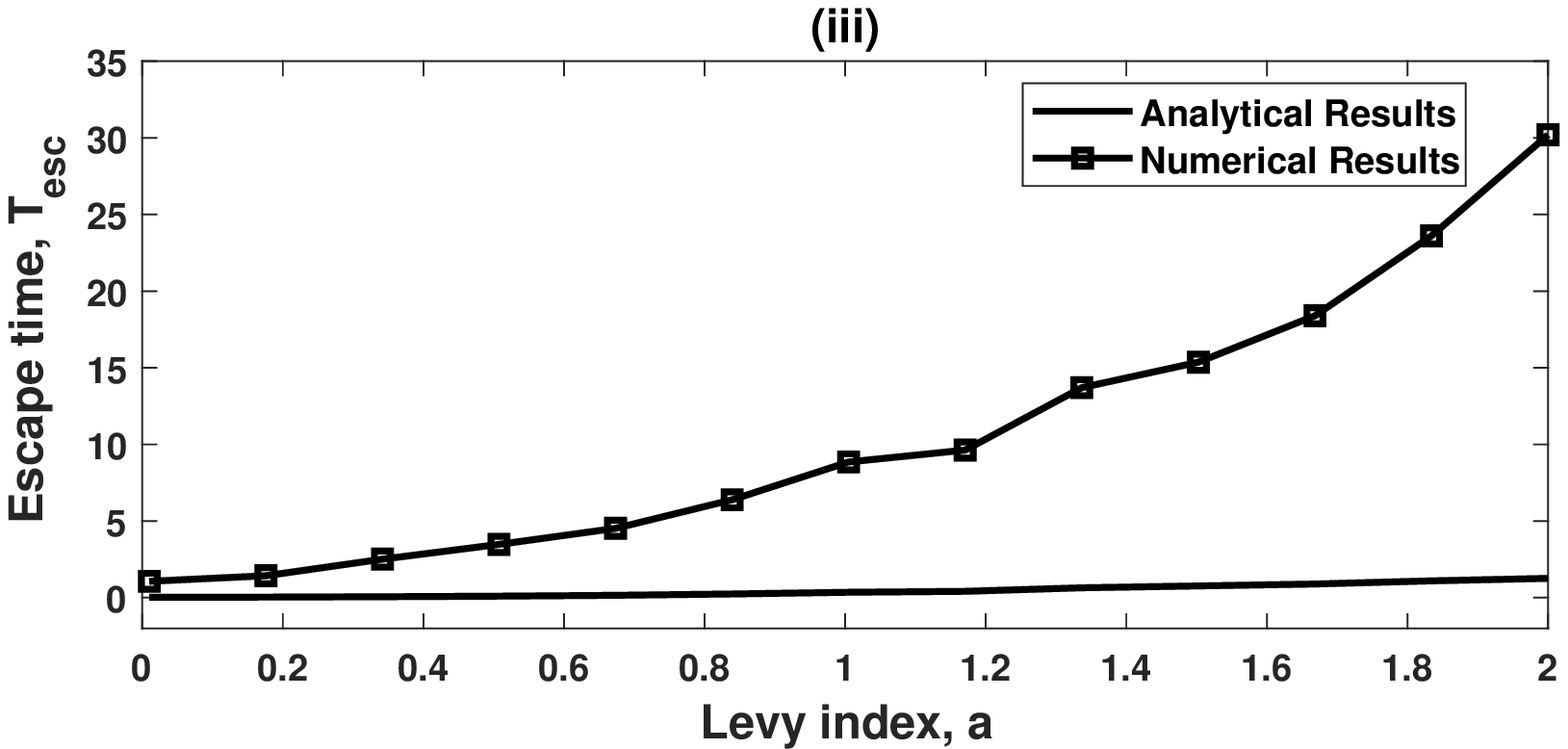}
\includegraphics[height=5cm,width=8.0cm]{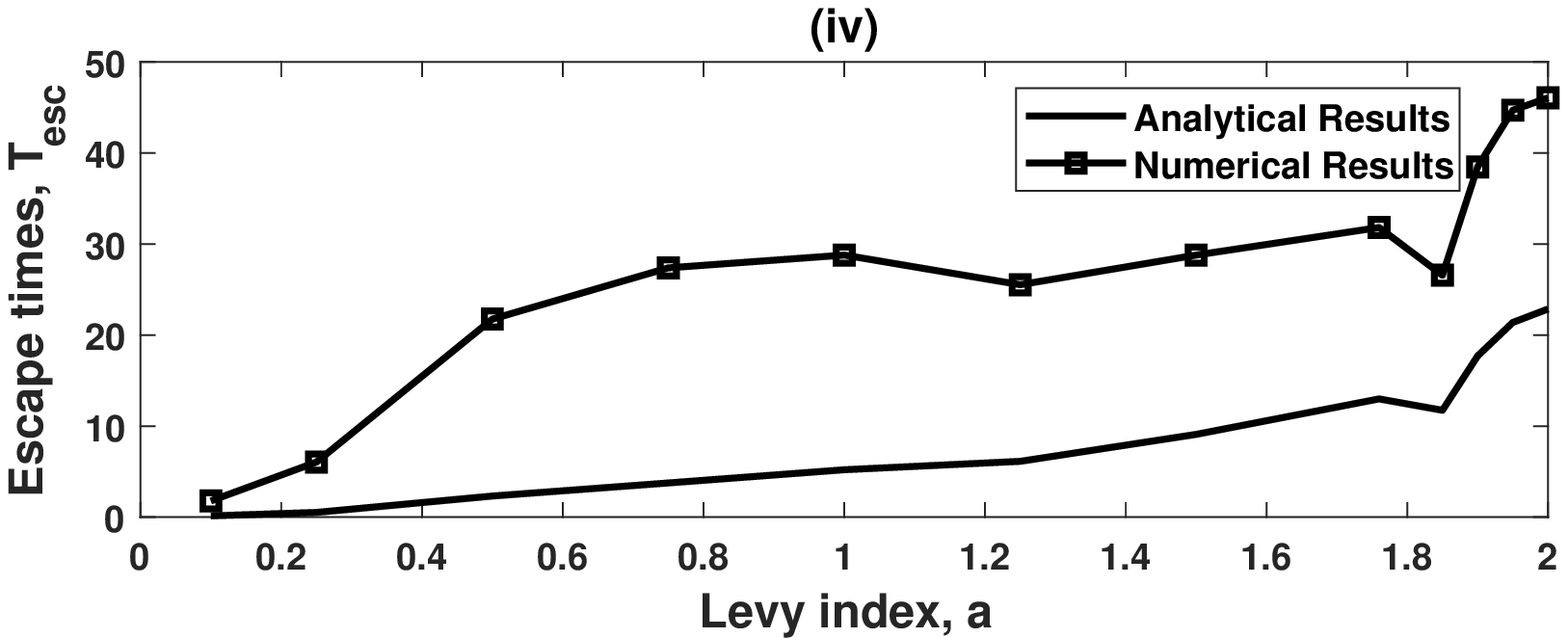}
\caption{\it Compaison between analytical and numerical results for
the asymetric quasi-potential (i); (ii), and  the symetric potential(iii),(iv).
(i,ii) $A_1\to A_3$; (ii,iv) $A_3\to A_1$ with $D=1.0$.}
\label{fig9}
\end{center}
\end{figure}

Numerical simulations can be used to show the effect of the Levy index $a$ on the pseudo-potential associated with  Eq. (\ref{tau_Levy_gen}), that is on the global stability properties of the attractors.
In other words, one can estimate from numerical simulations the average time that a
particle confined in the potential well spends to move to the other well.
The analytic approximations developed in the previous subsection can thus be checked against the numerical results.
A discrepancy is to be expected, for the theory of the L\'evy noise has not been fully extended to the pseudo-potentials.
 Fig. \ref{fig9} presents the comparison between the numerical studies and the analytical results, for the transitions in the two cases of symmetric and asymmetric pseudo-potentials.
It appears in the data that the comparison between the analytical and numerical results is acceptable for the L\'evy  $a=0.1$, but the agreement progressively deteriorates when the index $a$ increases.

\begin{table}
\begin{center}
\begin{tabular}{|c|c|c|}
  \hline
  $Levy\,\, index\,\,a$ & $\frac{\Delta U_1}{\Delta U_3}(AS_1)$ & $\frac{\Delta U_1}{\Delta U_3}(S_1)$ \\
  \hline
  0.1&	1.2&	1.22\\
0.25&	2.22&	1.57\\
0.5&	1.006	&0.94\\
0.75&	1.29	&1.24\\
1.0	& 1.2	& 0.921\\
1.25&	1.27&	0.85\\
1.5	&1.2& 0.84\\
1.75 &	1.58 &	1.58\\
1.85	&1.512&	1.068\\
1.9 & 2.34&	0.87\\
1.95	&3.9 &	0.67\\
1.99&7.26&0.617\\
2.0&3.23&0.58\\
  \hline
\end{tabular}
 \caption{\it{Dependencies of the ratio $\Delta U_1/\Delta U_3$ versus the L\'evy index $a$ for the
 asymmetric ($AS_1$) and symmetric ($S_1$) pseudo-potentials.}}
  \label{Tab5}
\end{center}
\end{table}

\subsection{Estimate of the effective energy barriers and residence times of the attractors}

To determine the effective energy barriers, $\Delta U_{1,3}$,
it is important to notice from the behavior of the type reported
in Fig. \ref{fig3} that there exist different regimes for the
dependence of the escape times $T_ {esc}$, depending on the L\'evy noise parameters.
For example, in the low noise intensity regime, the data in Fig. \ref{fig3a} and \ref{fig4a}
 exhibit an exponential behavior of the escape time $T_{esc}$ as a function of the inverse noise intensity, $1/D$.

{\bf GF: I am not sure that the figures as they are now are sufficient to demonstrate the behavior that is here described. Also, it is not clear the procedure, I have tried to add some details, please complete the description. Moreover, for L\'evy flights equation (\ref{Arrhenius}) does not hold, one should use (\ref{tau_Levy_gen}).  Finally, it might be interesting to check other quantities that are not $\Delta U$.}

\begin{figure}%[htbp]
\begin{center}
\includegraphics[height=6cm,width=8.0cm]{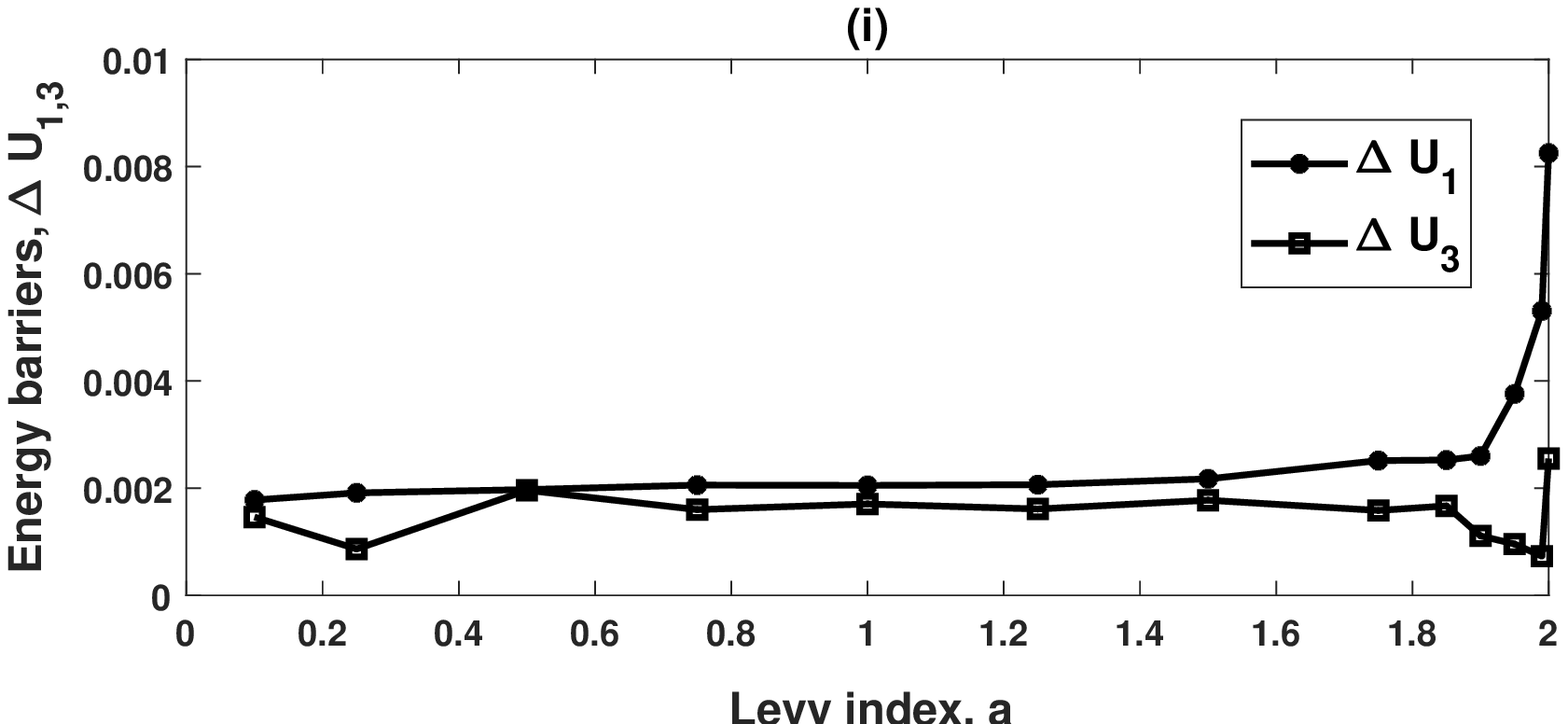}
\includegraphics[height=6cm,width=8.0cm]{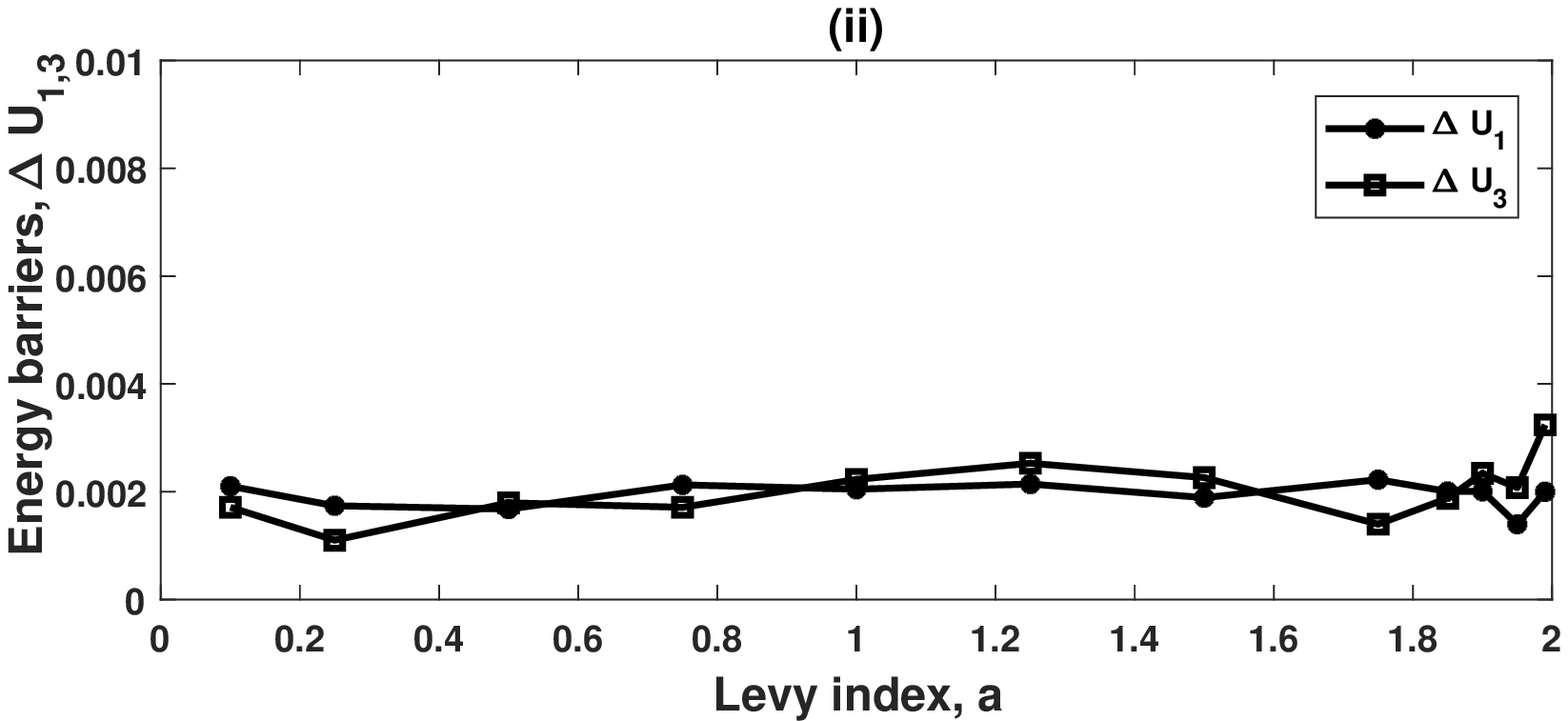}
\caption{\it Variation of the energy barriers  versus
the L\'evy index, $a$. Note that (i) correspond for the asymetric quasi-potential,
while  (ii) is for the symetric potential. The other parameter is $\mu=0.01$.}
\label{fig10}
\end{center}
\end{figure}

Fitting a straight line through the data points in the linear part
of Eq. (\ref{Arrhenius}) and measuring its slope, we obtain an
estimate of $\Delta U_1$ and $\Delta U_3$, the effective activation
energies for the escape from the limit-cycle attractors $A_1$ and $A_3$, respectively.
We have determined and shown in Figure \ref{fig10} the
dependencies of the energy barriers on the  Levy index $a$, for the escape transitions $A_1 \to A_3$
 and vise versa. We note that when we add the Levy noise term on Eq. (\ref{eq10}),
 the symmetric properties of the pseudo-potentials are less and less
 observed especially when the L\'evy index is less than $2$,
  but is observed well when the index is equal to $2$.
    Figure \ref{fig10} reveals that the two energy
 barriers are roughly equivalent when $a <2.0$.
  Figure \ref{fig10}(i) shows the variation in the effective energy
barriers versus the L\'evy index $a$  with the set of parameters
$AS_1$. The effective energy barriers increase slowly
when the L\'evy index $a$ increases,
and the behaviors strongly depend upon the L\'evy index.
For instance $\frac{\Delta U_1}{\Delta U_3}(a=0.1)\equiv 1.2$,
one concludes that the limit-cycle attractor with amplitude $A_1$ of the modified
van der Pol oscillator is much more stable than the limit cycle attractor with amplitude
$A_3$ (with respect to L\'evy noise).
The increase in the L\'evy index will not modify the properties
of the stability that we have just underlined, but the attractor
around the limit cycle with amplitude $A_1$ becomes more and more stable
when this index is close to $2$.
We will have gradually $\frac{\Delta U_1}{\Delta U_3}(a = 1.9) \equiv 2.34$,
 $\frac{\Delta U_1}{\Delta U_3}(a = 1.99) \equiv 7.3$
 and  $\frac{\Delta U_1}{\Delta U_3}(a = 2.0) \equiv 3.4$.
 Figure \ref{fig10}(ii) corresponds on the symmetric pseudo-potential and
 one shows the dependencies of the effective energy
barriers versus the L\'evy index $a$  with the set of parameters
$S_1$. The behaviors of the effective energy barriers increase slowly
with the increase of the L\'evy index $a$,
and depend upon the escape transitions.
It appears that when the L\'evy index $a$ takes the value $a=0.1$,
the limit cycle amplitude $A_1$ is more stable than the limit
cycle amplitude $A_3$, (\emph{i.e.} $\frac{\Delta U_1 }{\Delta U_3} \equiv 1.22$),
when the L\'evy index increases moreover, the same scenario
continues until the value $a <1.0$, in which the opposite
phenomenon occurs and the  limit cycle amplitude $A_3$ becomes
more stable, $\frac{\Delta U_1 }{ \Delta U_3} \equiv 0.9$. When the L\'evy index
more increases  until $a = 1.75$, we have a situation reversal
where the  limit cycle amplitude $A_1$ becomes more stable since
$\frac{\Delta U_1}{\Delta U_3}(a=1.75) \equiv 1.52$. When $a\ge 1.8$, it is the limit cycle amplitude $A_3$
 which returns stable that the limit cycle of amplitude $A_1$.
 This cascade of scenarios appears in table \ref{Tab5}
 where we have grouped the behavior of the ratio of the energy barrier values
 for the two escapes transitions, in the case of asymmetric and symmetric pseudo-potentials.
 In order to make an equivalence between the depth of a pseudo-potential
 well and the stability of the attractor associated with this pseudo-potential sink,
 we will evaluate the time spend by a confined particle will put around each potential well,
 \emph{i.e.} on each attractor.

\begin{figure}%[htbp]
\begin{center}
\includegraphics[height=6cm,width=8.0cm]{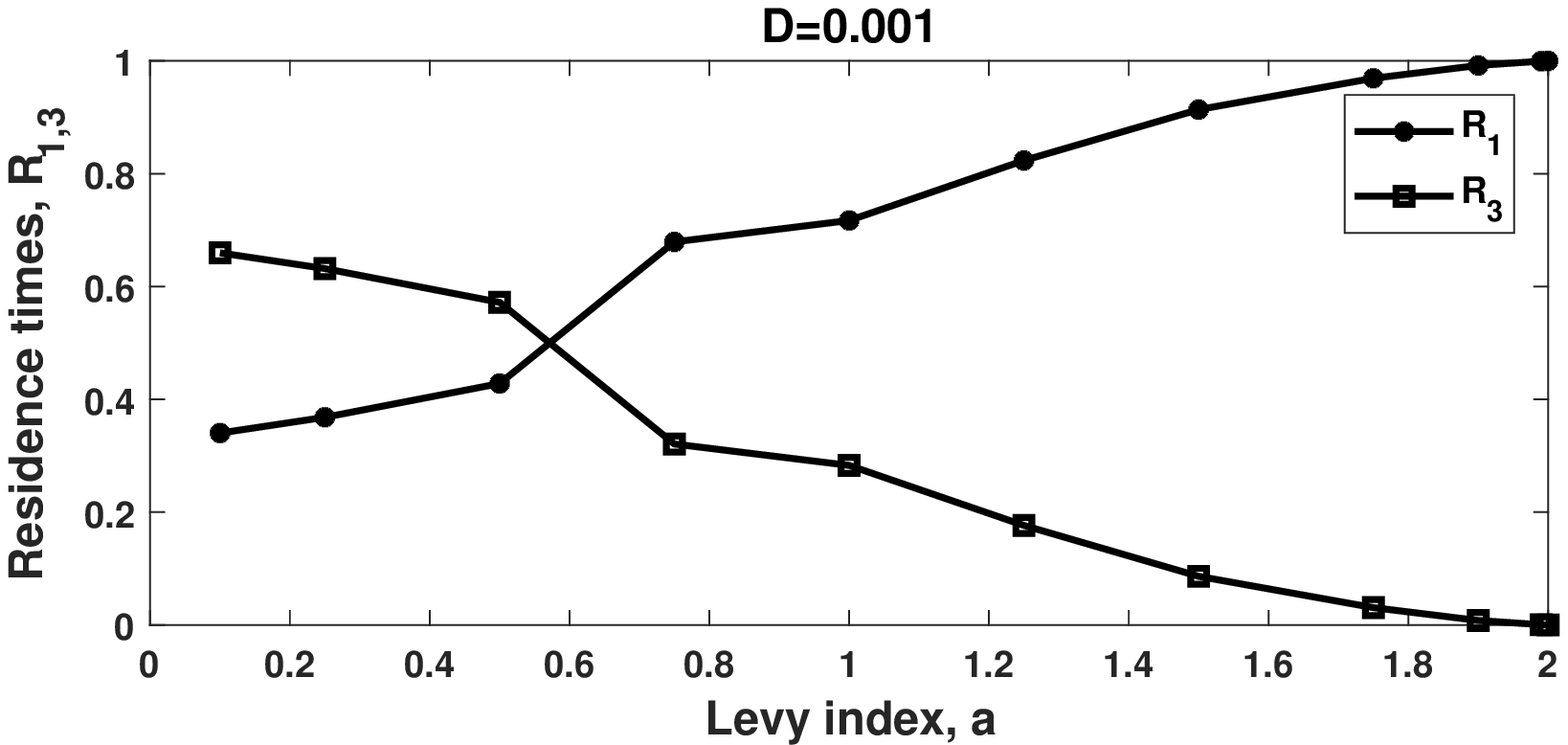}
\includegraphics[height=6cm,width=8.0cm]{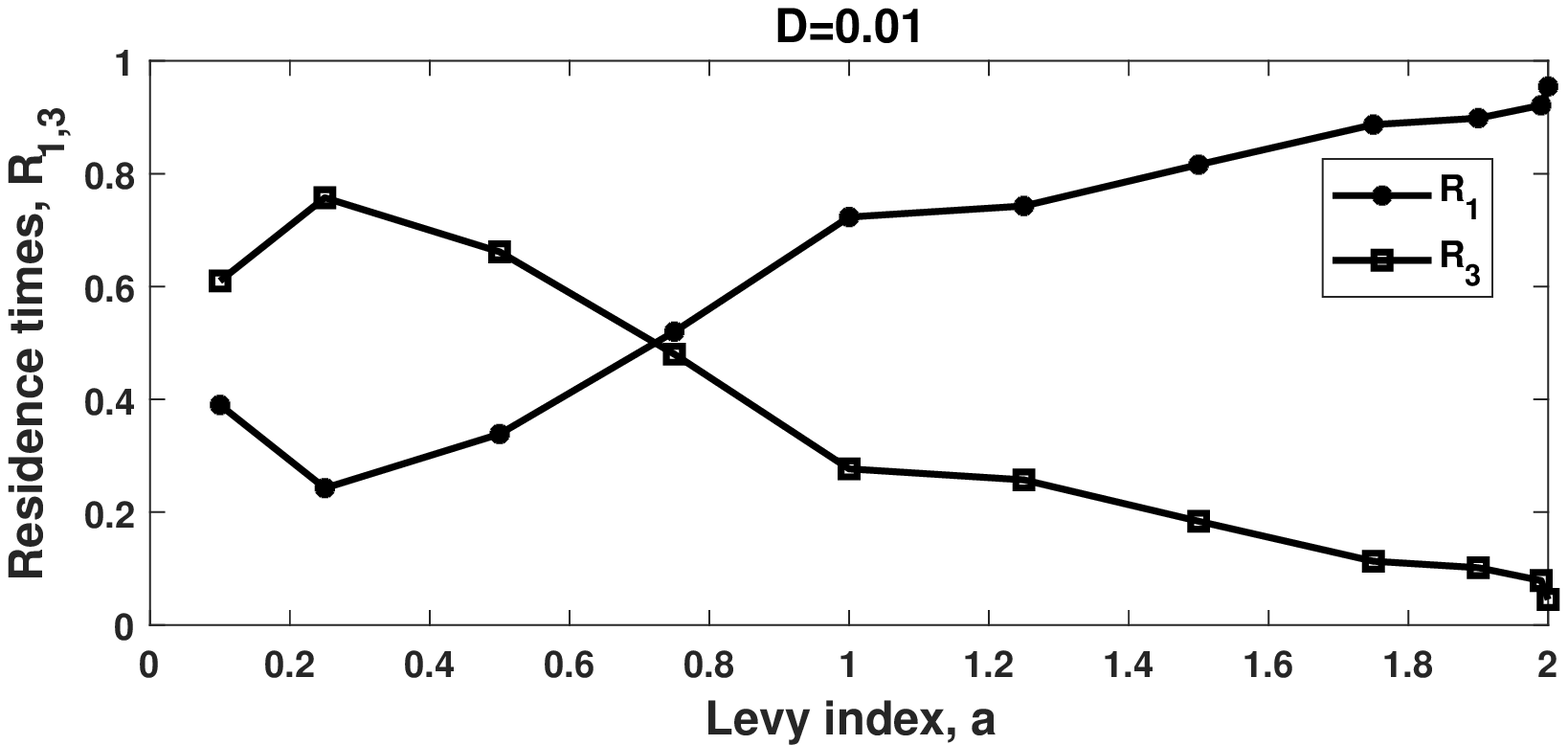}
\caption{\it
Residence times $R_{1,3}$ as a function of the L\'evy index, $a$ for different values
of the noise intensity, $D$,
 for the asymetric quasi-potential. The other parameter is $\mu=0.01$.}
\label{fig11}
\end{center}
\end{figure}

\begin{figure}%[htbp]
\begin{center}
\includegraphics[height=6cm,width=8.0cm]{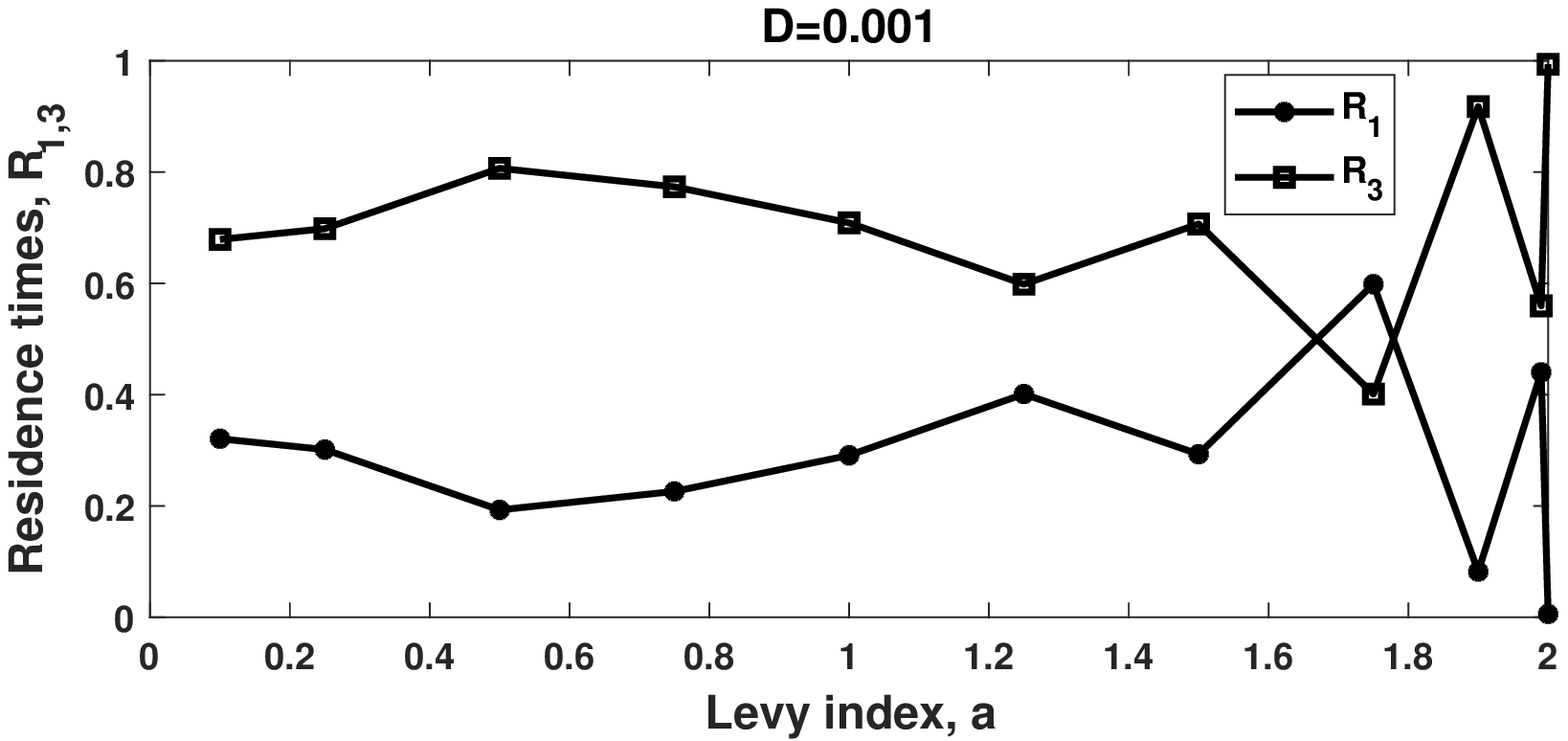}
\includegraphics[height=6cm,width=8.0cm]{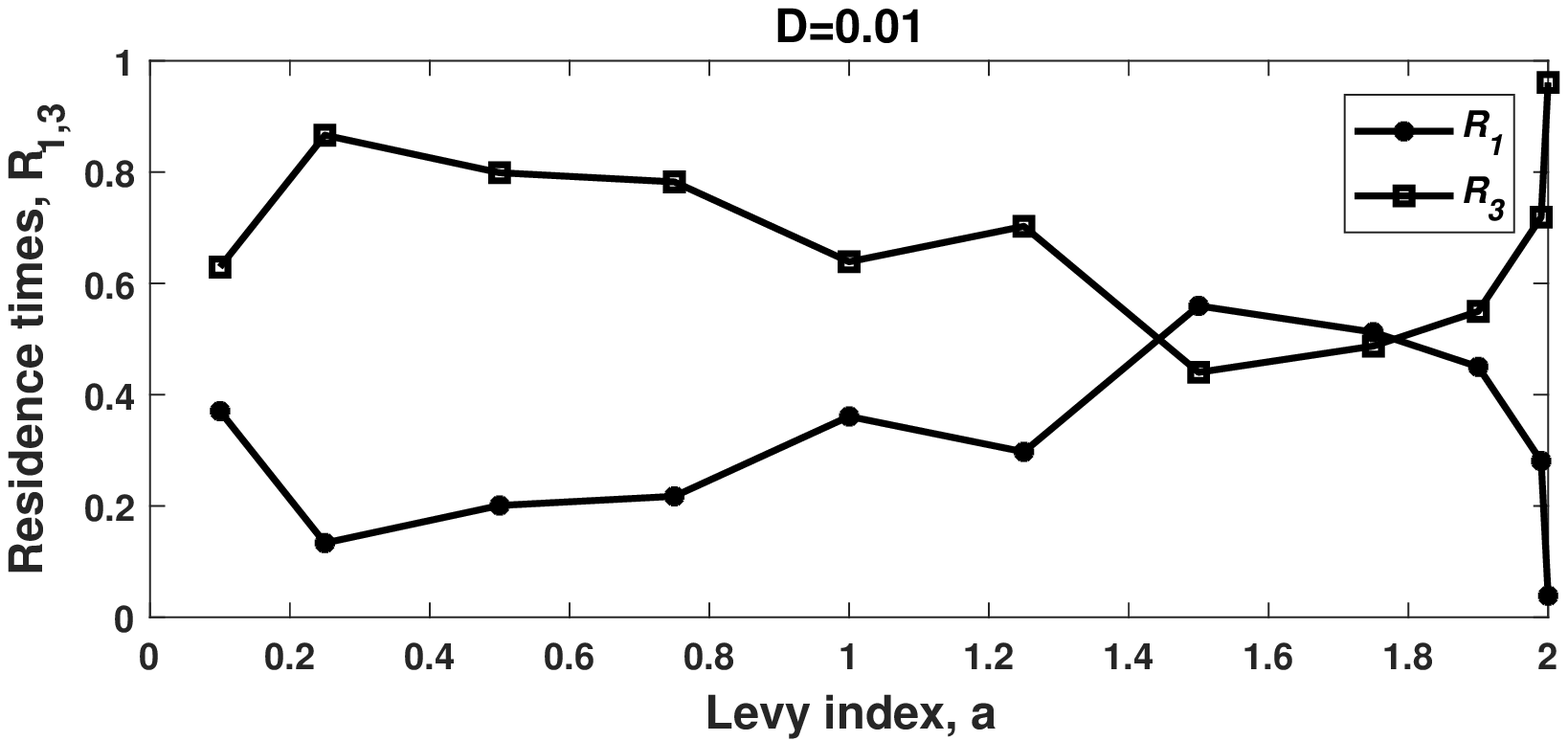}
\caption{\it
Residence times $R_{1,3}$ as a function of the L\'evy index, $a$ for different values
of the noise intensity, $D$,
 for the symetric quasi-potential. The other parameter is $\mu=0.01$.}
\label{fig12}
\end{center}
\end{figure}

Let us note  that “short” and “long” might be very
different. To measure the different properties, we compute
the average persistence or residence time $R_{1,3}$ on the
attractor with limit-cycle amplitude $A_{1,3}$ as
\begin{equation}\label{eq19}
  R_j=\frac{T_j}{T_1+T_3},\quad j=1,3
\end{equation}
where $T_{1,3}$ is the escape time from the first attractor $A_1$ (\emph{i.e.} $T_1=T_{esc}(A_1\to A_3)$) or
third attractor $A_3$ (\emph{i.e.} $T_3=T_{esc}(A_3\to A_1$).
Figures \ref{fig11} and \ref{fig12} show the effects of the noise intensity on the dependencies
the residence times $R_{1,3}$ as a function of the L\'evy index, $a$
for the asymmetric and the symmetric pseudo-potentials, respectively.
For the case of asymmetric pseudo-potential (\emph{i.e.} the parameters $AS_1$),  for
noise intensity around $D=1/ 1000$ and with the L\'evy index fix at $a=0.1$,
we get $R_3(a=0.1)=0.659$, and obviously
$R_1(a=0.1)=0.3402$, (see Fig. \ref{fig11}) \emph{i.e.}, the system will spend
$65.9\%$ of the time
on the third attractor $A_3$ and $34.02\%$ on the first attractor $A_1$.
When we increase Levy's index, the residence time on attractor $A_1$
increases while that on attractor $A_3$ decreases. It appears through Fig.\ref{fig11}(i)
 that at $a = 0.6$, the system will spend the same time on the two attractors $A_1$
  and $A_3$, \emph{i.e.} $R_1 (a=0.6)\equiv R_3(a=0.6)$. By further increasing the L\'evy index,
  $R_1$ continues to increase while $R_3$ continues to decrease, we will
  still have $ R_1> R_3 $ when $ a> 0.6.$
  By changing the intensity of the noise, \emph{i.e.} $D = 1/100$, we will obtain
  the same scenario as the one described previously, but with the difference
  that the system will spend the same time on the two attractors, (\emph{i.e.} $R_1 \equiv R_3$)
   when the L\'ey index, $ a $ is around $0.7$.
   Figure \ref{fig12} shows the variation of the residence times $R_{1,3}$
   as a function of the L\'evy index for the symmetric pseudo-potential, with two values of the noise intensity:
   $D=1/1000$ and $D=1/100$.
   We note that unlike the case of asymmetric pseudo-potential,
   the system will take more time on the attractor $A_3$ because $R_3> R_1$
    when the index of L\'evy increases. But we will note
    a small window of the L\'evy index $a$ where we observe
    $ R_3 <R_1, $ which can be neglected when the number of numerical
     iteration processes becomes very large.
%\section{Stochastic bifurcations}

%\subsection{Statement of the problem}

%\subsection{Stationary probability distribution}

%\subsection{Influence of Levy's noise parameters}

%\subsection{Comparison with numerical simulations}

\section{Conclusion}

We have considered through numerical simulations the effects of L\'evy noise on the birhythmic van der Pol system.
After presenting the self-sustained model used, we briefly recalled
the birhythmic properties on the free noise  model. We then give
the information about the L\'evy noise process and indicates the algorithm
we used to generate the L\'evy noise. The L\'evy probability density  function
was represented as a function of different L\'evy stability parameters,
showing its symmetric and asymmetric character.
To find the effects of L\'evy noise on the occurrence of large deviations, that
is excursions from an attractor to another, we have modified the
functional form of the \emph{Arrhenius}-like behavior, Eqs.(13) and
found that the escapes are governed, for the L\'evy noise systems by the law Eq.(14),
depends in addition to the noise intensity, $D$, to the L\'evy parameters.
Adding a random excitation, we have found that the system
crosses the boundary between the basins of attraction
(i.e., moves across the unstable limit cycle with amplitude
$A_2$). The mean time $T_{esc}$ to escape from one limit-cycle attractor
to the other has been estimated in the low-noise limit, and it
is proposed as a measure of the attractor’s global stability.
We have found that as in other
systems that exhibit noise induced switches between two attractors,
the escape times can be very different and significantly depend to the L\'evy index, $a$.
It appeared that, increasing the L\'evy index has the important influence on the
birhythmic properties and then on the stability analysis. For instance, the pseudo-potential
which initially was asymmetrical (symmetrical) becomes with the variation of the L\'evy index,
symmetric (asymmetrical). And therefore for the fixed value of the noise intensity,
the means escape time has increased, decreased or remains almost constant depending
on the shape of the resulting pseudo-potential well.
By considering the variation in the mean escape time $T_{esc}$ versus
the inverse noise intensity $1/D$, the slope of the linear
part has enabled us to summarize the results in the form of an
effective activation energy barrier, which is function of the
L\'evy index $a$.

We conclude that.............

\section*{Conflict of Interest}
The authors declare that they have no conflict of interest.

\section*{Acknowledgments}
R.Y. undertook this work with the support of the German Academic Exchange Service(DAAD), Germany. He acknowledges the support of the Potsdam Institute for Climate Impact Research (PIK), Potsdam, Germany.

\newpage
%\begin{thebibliography}{}

\newpage

\end{document}